\documentclass[fleqn,usenatbib]{mnras}

\usepackage[T1]{fontenc}

\DeclareRobustCommand{\VAN}[3]{#2}
\let\VANthebibliography\thebibliography
\def\thebibliography{\DeclareRobustCommand{\VAN}[3]{##3}\VANthebibliography}

\usepackage{graphicx}	
\usepackage{amsmath}	
\usepackage{amssymb}	
\usepackage{gensymb}

\title[CO emission shows inversion $\&$ hot spot in WASP-33b]{Carbon monoxide emission lines reveal an inverted atmosphere in the ultra hot Jupiter WASP-33 b consistent with an eastward hot spot}

\author[van Sluijs et al.]{
Lennart van Sluijs,$^{1,2}$\thanks{E-mail: lennart.vansluijs@physics.ox.ac.uk}
Jayne L. Birkby,$^{1,2,3}$
Joshua Lothringer,$^{4}$
Elspeth K. H. Lee,$^{5}$
Ian J. M. Crossfield,$^{6}$\newauthor
Vivien Parmentier,$^{7}$
Matteo Brogi,$^{8,9,10}$
Craig Kulesa,$^{11}$
Don McCarthy,$^{11}$
David Charbonneau$^{3}$
\\
$^{1}$Department of Astrophysics, University of Oxford, Oxford, OX1 3RH, United Kingdom\\
$^{2}$Anton Pannekoek Institute for Astronomy, University of Amsterdam, Amsterdam, 1098 XH, The Netherlands\\
$^{3}$Center for Astrophysics | Harvard $\&$ Smithsonian, 60 Garden Street, Cambridge, MA, 02138, USA\\
$^{4}$Department of Physics, Utah Valley University, Orem, UT 84058, USA\\
$^{5}$Center for Space and Habitability, University of Bern, Gesellschaftsstrasse 6, CH-3012 Bern, Switzerland\\
$^{6}$Department of Physics and Astronomy, University of Kansas, Lawrence, KS 66045, USA\\
$^{7}$Atmospheric, Oceanic $\&$ Planetary Physics, Department of Physics, University of Oxford, Oxford OX1 3PU, UK\\
$^{8}$Department of Physics, University of Warwick, Coventry CV4 7AL, UK\\
$^{9}$INAF-Osservatorio Astrofisico di Torino, Via Osservatorio 20, I-10025, Pino Torinese, Italy\\
$^{10}$Dipartimento di Fisica, Universit\`a degli Studi di Torino, via Pietro Giuria 1, I-10125, Torino, Italy \\
$^{11}$Steward Observatory, University of Arizona, Tucson, AZ, USA\\
}

\date{Accepted XXX. Received YYY; in original form ZZZ}

\pubyear{2022}

\usepackage{newtxtext,newtxmath}

\begin{document}
\label{firstpage}
\pagerange{\pageref{firstpage}--\pageref{lastpage}}
\maketitle

\begin{abstract}
We report the first detection of CO emission at high spectral resolution in the day-side infrared thermal spectrum of an exoplanet. These emission lines, found in the atmosphere of the transiting ultra hot Jupiter (UHJ) WASP-33 b, provide unambiguous evidence of its thermal inversion. Using spectra from the MMT Exoplanet Atmosphere Survey (MEASURE, $R\sim15,000$), covering pre- and post-eclipse phases, we cross-correlate with 1D PHOENIX spectral templates to detect CO at S/N=7.9 ($v_{\rm{sys}}=0.15^{+0.64}_{-0.65}$ km/s, $K_{\rm{p}}=229.5^{+1.1}_{-1.0}$ km/s). Moreover, using Cross-Correlation-to-log-Likelihood mapping, we find that the scaling parameter which controls the spectral line contrast changes with phase. We thus use the general circulation model SPARC/MITgcm post-processed by the 3D gCMCRT radiative transfer code to interpret this variation, finding it consistent with an eastward-shifted hot spot. Pre-eclipse, when the hot spot faces Earth, the thermal profiles are shallower leading to smaller line contrast despite greater overall flux. Post-eclipse, the western part of the day-side faces Earth and has much steeper thermal profiles, leading to larger line contrast despite less overall flux. This demonstrates that within the log-likelihood framework, even relatively moderate resolution spectra can be used to understand the 3D nature of close-in exoplanets, and that resolution can be traded for photon-collecting power when the induced Doppler-shift is sufficiently large. We highlight CO as a good probe of UHJ thermal structure and dynamics that does not suffer from stellar activity, unlike species that are also present in the host star e.g. iron lines.
\end{abstract}



\section{Introduction}
Ultra hot Jupiters (UHJs) provide the unique opportunity to study gaseous planets in a strongly irradiated environment to learn about their composition and global circulation patterns \citep[e.g. for a review see][]{Showman2020, Fortney2021}. Their high day-side temperatures above 2200 K provide a unique window to directly detect volatile species in their vapor phase \citep[e.g.][]{Visscher2010, Parmentier2018, Hoeijmakers2018, Lothringer2018, Kitzmann2018, Hoeijmakers2019, Ehrenreich2020, Merritt2021, Cont2021}. Due to their tidally locked rotation, they have an extreme day-to-night-side temperature contrast \citep[e.g.][]{Knutson2007}. Consequently, global circulation models (GCMs) predict strong global jets and hot spots \citep{Showman2009, Menou2009, Parmentier2018, Tan2019}. This combined with their relatively high star-to-planet contrast ratio, makes them ideal targets for atmospheric characterisation with current observing facilities \citep[e.g.][]{Kreidberg2018b, Birkby2018}.
Hydrodynamic simulations and theoretical calculations predict that hot spots occur in ultra hot Jupiters with an eastward offset from the sub-stellar point due to global wind circulation patterns \citep[e.g.][]{Showman2002,Dobbs-Dixon2008,Menou2009,Dobbs-Dixon2010,Rauscher2010,Showman2011,Perez-Becker2013,Debras2020}. Most observations support this eastward offset hot spot prediction \citep{Harrington2006, Cowan2007, Knutson2007, Knutson2009, Charbonneau2008, Swain2009, Crossfield2010, Wong2016}, but for some UHJs westward offsets have been observed as well \citep{Armstrong2016, Dang2018, Jackson2019, Bell2019, vonEssen2020, Herman2022}. Several mechanisms including cloud asymmetries, asynchronous rotation and magnetohydrodynamical effects have been proposed to explain these westward offset hot spots \citep[e.g.][]{Hindle2021}.
    Thermal inversions of the pressure-temperature profiles (P-T profiles) of UHJs were at first predicted due to the the strong optical/UV absorbing molecules TiO and VO \citep[e.g.][]{Hubeny2003}. Later work found absorption by other atomic and molecular species, in addition to TiO and VO, can also result in thermal inversions \citep[][]{Lothringer2018, Arcangeli2018, Gandhi2019}. This is due to a combination of short-wavelength stellar irradiation around early type stars and absorption of these wavelengths by continuous opacity sources, metal atoms, metal hydrides, metal oxides, SiO, or possibly even disequilibrium species like SH \citep[][]{Evans2018}. Analysis of Spitzer and Hubble Space Telescope (HST) secondary eclipse observations support this scenario as they detect emission signatures suggestive of an inverted stratosphere for a handful of UHJs
\citep[e.g.][]{Deming2012,
Haynes2015,
Evans2016, Evans2017,
Sheppard2017,
Arcangeli2018,
Kreidberg2018,
Mansfield2018,
Nugroho2020a,
Yan2020,
Garhart2020, Baxter2021}.
One well-studied UHJ for which a thermal inversion was predicted by atmospheric modeling is WASP-33 b. This planet orbits every 1.22 days around an A5 star  \citep{Cameron2010}. The first observational evidence of a thermal inversion of WASP-33 b's atmospheric profile was excess infrared emission observed by Spitzer from secondary eclipse observations and \citep{Deming2012, Zhang2018} and NIR low-resolution HST/Wide Field Camera 3 spectra, possibly due to TiO \citep{Haynes2015}. Furthermore, \citep{Zhang2018} find a phase angle offset by $-12.8 \pm 5.8 \degree$ in the Spitzer 3.6$\micron$-band, consistent with an eastward hospot. On the contrary, \citep{vonEssen2020} find a $28.7 \pm 7.1 \degree$ phase angle offset in the optical with TESS, suggesting a westward offset hot spot instead, although they mention host star variability may introduce a spurious westward offset. Recently, \citet{Herman2022} report day-to-night brightness contrast variations and a $22 \pm 12 \degree$ westward phase offset from their detection of Fe-I emission lines.
For these low resolution observations, stellar pulsations of the $\delta$ Scuti pulsating host star induce variability of the stellar continuum \citep{Herrero2011}, limiting observational constraints as they induce a quasi-sinusoidal trend in transit light curve observations \citep[e.g.][]{Deming2012, deMooij2013, Garhart2020, vonEssen2019, vonEssen2020}.
Furthermore, stellar pulsations also affect high resolution spectral observations targeting any molecule present in both the stellar and exoplanetary atmosphere due to changes in the spectral line shape \citep{Herrero2011}. For WASP-33 b this has been observed for the detection of neutral iron lines by \citet{Nugroho2020, Herman2022}.
Pursuits to follow-up the tentative low-resolution detection of TiO in the atmosphere of WASP-33 b using high resolution spectra have lead to inconclusive results so far. \cite{Nugroho2017} detected the the optical TiO molecular signature using the High Dispersion Spectrograph on the Subaru Telescope. This detection was later challenged by a re-analysis of the same data by \cite{Serindag2021} who unexpectedly found a slightly weaker signal using the updated TiO ExoMol TOTO line list. Furthermore, \cite{Herman2020} analysed emission and transmission spectra from the Echelle SpectroPolarimetric Device for the Observation of Stars (ESPaDOnS) on the Canada-France-Hawaii Telescope and the High Resolution Echelle Spectrometer (HIRES) on the Keck telescope, but are unable corroborate their observations with a thermal inversion due to TiO. There is thus a lack of a current consensus regarding the observational evidence for TiO in the atmosphere of WASP-33 b.
Despite the elusive results for TiO, several new detections suggest WASP-33 b must have a thermally inverted atmosphere. Recently, emission signatures of FeI \citep{Nugroho2020, Cont2021}, neutral Si \citep{Cont2021b} and OH \citep{Nugroho2021} have been detected. \citet{Nugroho2021} also marginally detect $\textrm{H}_2\textrm{O}$, consistent with the theoretical predictions that $\rm{H}_{2}O$ molecules dissociate into OH and $\textrm{H}^-$ at the temperature regime of UHJs. Other atmospheric detections for WASP-33 b include Ca II \citep{Yan2019}, the hydrogen H$\alpha$, H$\beta$ and H$\gamma$ Balmer lines \citep{Cauley2021, Yan2021, Borsa2021}, evidence for AlO \citep{vonEssen2019}, and lastly, \citet{Kesseli2020} place upper limits on the Volume Mixing Ratio (VMR) for FeH based on their null detection. These detections are thus in line with predictions by \citet{Lothringer2018, Gandhi2019} of thermally inverted UHJ atmospheres, regardless of the presence of TiO or VO.
%

%
In this paper we report CO emission lines in the atmosphere of WASP-33 b using observations with the MMT in Arizona, USA, equipped with the ARizona Infrared imager and Echelle Spectrograph (ARIES). It is the first reported detection of CO emission lines detected with high resolution cross correlation spectroscopy (HRCCS). During the refereeing process of this paper, \citet{Yan2022} independently reported a detection of CO emission lines in the atmosphere of WASP-33 b using GIANO-B, and CO emission lines are also subsequently reported in the atmospheres of WASP-121 b \citet{Holmberg2022} and WASP-18 b \citet{Brogi2022}. However, we note that our detection here, which uses the large Doppler-shift induced by the planet's orbital motion, is at lower spectral resolution compared to other HRCCS observations, and still reveals information about the planet's atmospheric dynamics. The observations and the subsequent data reduction are discussed in Section~\ref{sec:obs}. Our atmospheric models are described in Section~\ref{sec:modeling}. These models are used to characterise the exoplanet's atmosphere using HRCCS as described in Section~\ref{sec:hrccs}. The results are presented in Section~\ref{sec:results} and discussed in Section~\ref{sec:discussion}. Our main conclusions and recommendations for future work are summarised in Section~\ref{sec:conclusions}.
\section{Observations and data processing}
\label{sec:obs}
\subsection{Observations}
\label{sec:observations}
\begin{figure}
    \centering
    \includegraphics[width=\columnwidth]{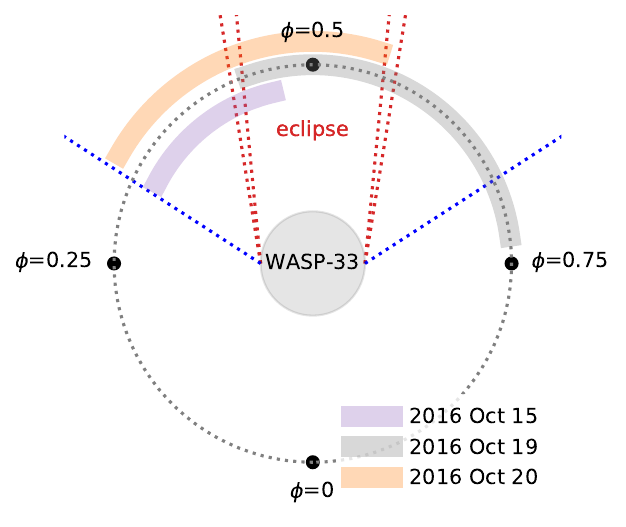}
    \caption{Orbital phase ($\phi$) coverage for each observing night of the WASP-33 data set. As can be seen, all nights contain some spectra taken during full eclipse of WASP-33 b. The blue dotted line indicates up to which phases we have symmetric phase coverage for both the pre- and post eclipse data.}
    \label{fig:phase_coverage}
\end{figure}

We observed the ultra hot Jupiter WASP-33 b with the ARizona Infrared imager and Echelle Spectrograph (ARIES; \citealt{McCarthy1998,Sarlot1999}) in combination with the f/15 adaptive secondary mirror and adaptive optics system at the 6.5-m MMT Observatory on Mt Hopkins in Arizona, USA.
The adaptive secondary mirror provides a low thermal background, while augmenting the total throughput of the instrument.  ARIES can provide spectral observations using both long slit and echelle spectrograph modes. It can observe in the 1-5 $\micron$ range and at spectral resolutions of 2,000-30,000. The observations of WASP-33 b presented in this work are the first of a larger survey called MEASURE (MMT Exoplanet Atmosphere Survey) which contains observations of a diverse set of eleven exoplanets with a wide range of temperatures, masses and radii.
For these observations of WASP-33 b, we obtained echelle spectra in the 1.37-2.56 $\micron$ wavelength range during three half nights in October 2016 using the ARIES/MMT  combination. In total 211 spectra were obtained with an exposure time of $300$ s per frame at a theoretical instrumental resolution of $R = 30,000$ using the $1^{\prime\prime}\times0.2^{\prime\prime}$ slit and the f/5.6 camera mode. The simultaneously operated ARIES imager was used for in-slit guiding, with occasional manual offsets made to correct any sustained drift during an exposure not corrected by the AO guiding system. 
An overview of our observed orbital phase coverage\footnote{Orbital phases have been calculated using the ephemerides in Table~\ref{tab:wasp33_system}, which contains all relevant WASP-33 system parameters used throughout this work.} is shown in Fig.~\ref{fig:phase_coverage} and our observational parameters and data quality are summarised in Table~\ref{tab:obs}. The layout of the 26 echelle orders of ARIES is shown in the top-left panel of Fig.~\ref{fig:fringecorr}. The three observing nights cover both pre- and post-eclipse orbital phases, but the post-eclipse data covers significantly more of the orbit, almost to quadrature, where the planet shows half its day-side and half its night-side to the Earth.
At the beginning and end of each half night, we observed a set of calibration images. This included a set of dark frames obtained using a blank in the filter wheel and the ARIES entrance covered to prevent any light entering the instrument. We observed flat-fields with the grating in, using an incandescent light bulb arranged such that its light reflected off an aluminumized board into the dichroic of ARIES. We also observed a thorium-argon lamp in a similar manner at the start of the nights to focus the spectrograph, but used the observed telluric absorption lines for simultaneous wavelength calibration.

\begin{table}
    \centering
    \begin{tabular}{cccccc}
    \hline
         Date&Total&No.&Phase&S/N$^{*}$  \\
             (UTC)&exposure &spectra&Coverage&\\
             &time (min)&&\\
    \hline\hline
         2016 Oct 15&220&44&0.331-0.469&91\\
         2016 Oct 19&465&93&0.461-0.731&26\\
         2016 Oct 20&370&74&0.337-0.554&71\\
         \hline
    \end{tabular}\\
    {$^{*}$This quantity refers to the median S/N per wavelength bin \\
    as measured from the extracted and normalised spectral time series\\
    of the 23rd spectral order around the $2.3 \micron$ CO emission region.}
    \caption{Observational parameters of the WASP-33 data set for each night. }
    \label{tab:obs}
\end{table}

\subsection{Data reduction: pre-processing}
\label{sec:pre-processing}
\begin{figure}
    \centering
    \includegraphics[width=\columnwidth]{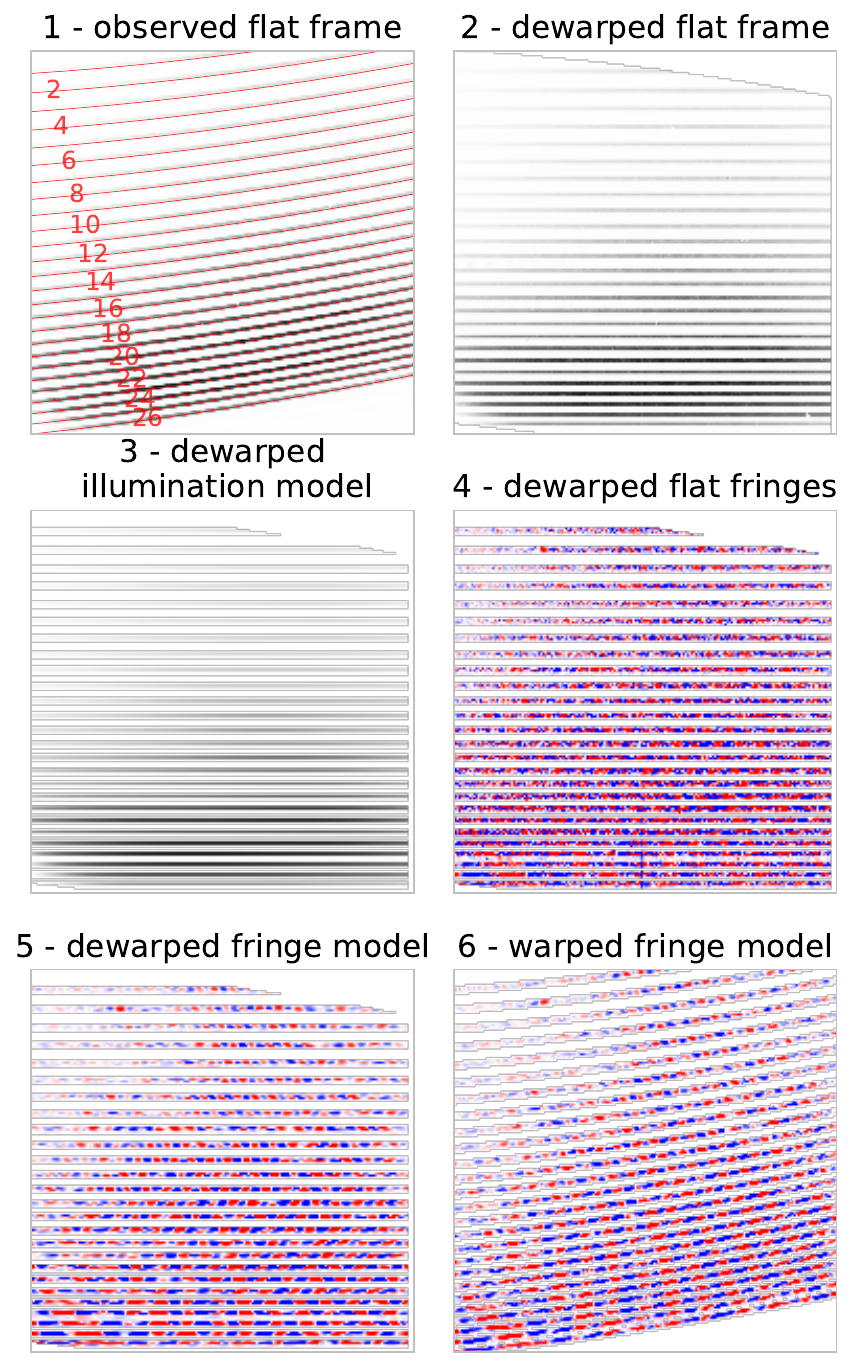}
    \caption{Flat fringe correction illustrated in six steps: (1) the observed flat frame with the fitted echelle traces overlaid in red (numbered from blue-to-red). (2) the flat frame after applying a 2D polynomial transform to straighten the echelle traces. (3) the 4th order polynomial row-by-row illumination model. (4) fringes in the dewarped flat frame. (5) row-by-row low-pass filter fringe model. (6) fringe model after applying inverse 2D polynomial transform.}
    \label{fig:fringecorr}
\end{figure}

Since this is the first time the MMT/ARIES combination has been used for exoplanet atmosphere characterisation, we developed a purpose-built end-to-end data reduction pipeline that we describe here first. This new pre-processing pipeline is publicly available on the author's GitHub\footnote{\url{https://github.com/lennartvansluijs/ARIES}}. As ARIES is currently being upgraded into MAPS \citep{Morzinski2020}, the instrument is no longer available in this exact form, but our data reduction pipeline uses a generalized approach which may be useful to reduction of the instrument when it returns. The procedure is described below, and is applied in the same way to all three nights of data:
\begin{enumerate}
\item Cross-talk correction: ARIES experiences cross-talk between the four quadrants of its detector which needs to be corrected. A bright pixel in one quadrant produces distinct, cross-shaped, negative shadows at the same position in all other three quadrants on the detector. This effect is apparent in all dark, flat, science frames.
We use the C-based {\sc corquad} routine provided by the ARIES team\footnote{See \url{http://66.194.178.32/~rfinn/pisces.html} to correct for the cross-talk. A Python-based version is available at: \url{https://github.com/jordan-stone/ARIES }.}. Three input parameters define the convolution kernel size for each detector quadrant. The parameter space is explored using a grid-based search to find the best cross-talk convolution kernels, which we define as those that minimize the standard deviation in a $10 \times 10$ box around a visually identified prominent shadow feature. We apply the {\sc corquad} routine using the best convolution kernels parameters to all dark, flat and science frames, before any further data reduction.
\item{Dark correction: we need to remove hot pixels from individual darks and combine them into a master dark. We identify hot pixels as $\geq 3 \sigma$-outliers. Like the shadow cross-talk features, they are cross-shaped and we replace them by the median of their neighbours outside of a cross-shaped footing in a $11 \times 11$ box around each hot pixel. We median combine all frames of identical exposure time into a master dark. We subtract the appropriate master dark from each science and flat frame.}
\item{Flat correction: all flats contain fringes arising from Fabry-P\'erot interference between the optical elements of the instrument. Fringes in the flat field introduce modulation in the continuum that would be erroneously divided into the science flat during flat field correction, reducing the sensitivity of our later procedures. To remove the fringes from the flats, we developed a modified version of the flat fringe correction by \cite{Stone2014}. The procedure is shown in Fig.~\ref{fig:fringecorr}. We use the \textit{getthem} routine from the publicly available Python package {\sc CERES}\footnote{\url{https://github.com/rabrahm/ceres}} \citep{Brahm2017} to fit the echelle traces as 4th order polynomials using a 10 pixel wide aperture. We use the solution of all 26 spectral orders to calculate a 2D polynomial transform to dewarp the flat frame. We fit each row by a 7th order polynomial to create a dewarped illumination model, essentially representing the dewarped blaze function. This model is subtracted from the dewarped flat frame to reveal the dewarped flat fringes. Most fringes have frequencies $< 0.025 \ \rm{pixel}^{-1}$. We apply a 6th order Butterworth high-pass filter with a cutoff-frequency $f_{\rm{c}} = 0.025 \ \rm{pixel}^{-1}$ to create a dewarped fringe model. We then use the inverse 2D polynomial transform to warp the fringes model. We remove the fringes from the observed flat frame by subtracting the warped fringes model. It is key this subtraction is done in the observed coordinate frame such that the observed flat was not interpolated by a transform operation, thus preservering any intra-pixel systematics.

After defringing all flats, we perform a hot pixel correction similar to the one done for the dark frames and create a master flat. The ARIES detector has a central column of dead pixels. We correct this column by replacing it by the median of the two neighbouring columns in the master flat and science frames. To create a master flat hot pixel map, we first normalize the master flat by dividing the illumination model and identifying pixels with values outside of the [0.5, 1.5] range, as hot pixels. We combine the master flat hot pixel map with the master dark hot pixel map and replace all master flat hot pixels by the fit from the illumination model. For each science frame we correct each flagged hot pixel by their closest neighbours in the dispersion direction of the spectrograph. Finally, the master flat is divided out of all of the science frames. The science frames still contain fringes at this stage, but these will be removed by applying a high-pass-filter to the extracted spectral time series in the post-processing stage of the pipeline.
\item Spectral order extraction: after the flat correction we can extract the spectral orders from the detector images. We cannot directly fit the detector echelle traces for the science frames as these traces are often disrupted due to telluric absorption lines. Instead we fit the echelle traces using the {\sc CERES} {\sc retrace} routine. This treats the initial set of master flat traces as an aperture mask and cross-correlates it with each science frame to determine the drift in the cross-dispersion direction. We use the {\sc CERES} {\sc Marsh}-algorithm \citep{Marsh1989} routine with a 5 pixel aperture radius and clipping $>$5$\sigma$ outliers. The final output of the pre-processing part of the data reduction pipeline is a spectral time series for each of the 26 spectral orders for each of the 3 observing nights of the system (see example for one order in the top panel of Fig.~\ref{fig:postprocessing}).
}
\end{enumerate}
\subsection{Data reduction: post-processing}
\label{sec:post-processing}
\begin{figure}
    \centering
    \includegraphics[width=\columnwidth]{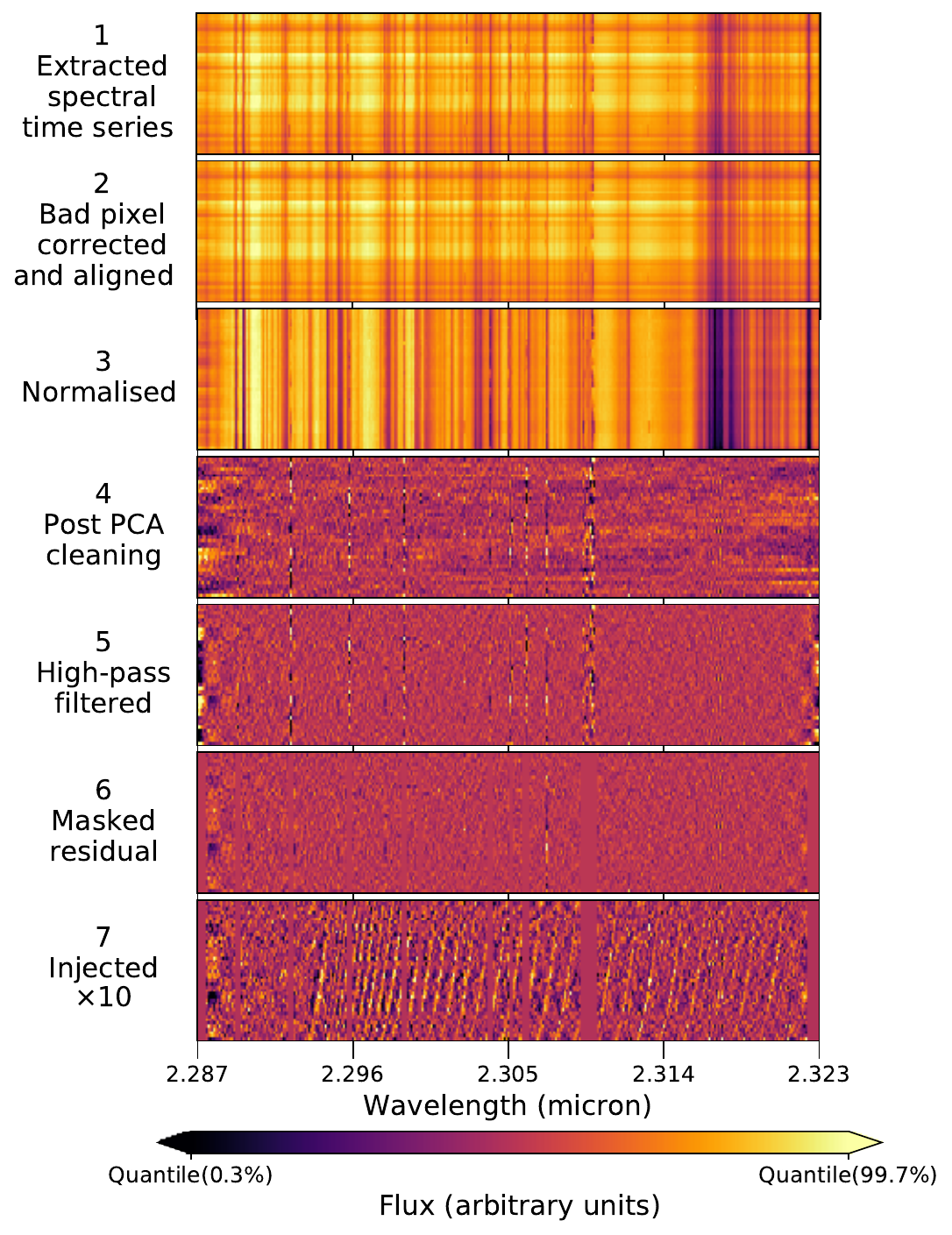}
    \caption{Example of post-processing procedure of a spectral time series for order 23 for the first observing night. \textit{First panel:} The extracted spectral time series from the processed detector image is shown in the top panel. \textit{Second panel:} after bad pixel/column correction and alignment with respect to the first spectrum. \textit{Third panel:} normalised by the median of the 50 brightest pixels per spectrum. \textit{Fourth panel:} after removing the first 7 components using Principal Component Analysis. \textit{Fifth panel:} After applying a high-pass-filter with cutoff-frequency of 0.02 $\mathrm{pixel}^{-1}$. \textit{Sixth panel:} After masking the columns with the worst residual structure. \textit{Seventh panel:} same as the previous panel, but with a  $\times$10 PHOENIX self-consistent model injected into the spectral time series between panels 2 and 3.}
    \label{fig:postprocessing}
\end{figure}

From the spectral time series we would like to extract the planet spectrum, but it is faint and buried in the noise. Thus we first need to remove the telluric and stellar lines. An overview of all steps of the post-processing procedure for a single spectral order covering the most line dense part of the CO spectrum is shown in Fig.~\ref{fig:postprocessing}. We perform the following steps to post-process the extracted spectral time series for each spectral order in a homogeneous manner:
\begin{enumerate}
    \item Bad pixel/column correction (see Fig.~\ref{fig:postprocessing}, panel 2): in each spectral time series, some bad pixel columns or detector artifacts may still be present in the spectral time series. To reveal these anomalies, we normalise each spectra by the median of each row, followed by subtracting the mean of each columns. 
    We use a Blob Detection Algorithm (an edge-detection algorithm), which uses a Laplacian of the Gaussian Filter to detect outliers, as described in more detail by \citet{Kong2013}. In our application we identify features of the size of a single pixel. These outliers are corrected by interpolating in the dispersion direction over the closest nearby good pixels. We ignore any convolution boundary effects here as they only affect the edge pixels and these will be clipped during the post-processing stage of the pipeline.
    Bad columns in the spectral time series are identified by taking the median of each spectral channel and applying a median-filter using a three pixel width and subtract this from the original. As before we use the Blob Detection Algorithm to identify 5$\sigma$-outliers in the residual time series which are considered bad columns and are corrected by interpolation from their nearest neighbouring good columns. We iteratively perform the bad pixel and column correction procedure five times to ensure all outliers are corrected. In total only $\sim$0.2\% of all pixels are identified as outliers.
    
    \item Alignment of spectra (see Fig.~\ref{fig:postprocessing}, panel 2): effects such as instrumental flexure, variation of the air mass or misalignment of the star on the spectrograph slit, can all cause sub-pixel drift of the spectra across the detector. We need to align the spectra to correct for these effects, else they will leave residuals later in the telluric line removal procedure. We did not fit a telluric model to each spectrum directly, as this adds the complication that the telluric model needs to match the observing conditions for each frame well. Instead we use the data themselves as a reference: the first spectrum of every night is used as a reference spectrum, thus the alignment will be in the telluric rest frame of the first spectrum. The alignment procedure is done in three steps. Firstly, the reference spectrum is cross-correlated with each frame. Secondly, we fit a Gaussian to the cross-correlation function (CCF) to determine the drift on a sub-pixel level for each frame. Finally, we then shift each spectra using linear interpolation to align them with the reference spectra.
    
    \item Telluric wavelength calibration: after alignment, we perform wavelength calibration using the observed telluric lines. We generate a telluric model using {\sc ATRAN}\footnote{\url{https://atran.arc.nasa.gov/cgi-bin/atran/atran.cgi}} \citep{Lor92}. For each aligned spectral order we cross-match by-eye telluric lines in the model to the corresponding telluric lines in the median of our observed spectra. We robustly fit a polynomial to all these points for each order to obtain a wavelength solution. Robust here refers to the fact that we iteratively clip $\geq 5 \sigma$ outliers and refitting a polynomial until no more outliers are found. By default we fit 3rd order polynomials, but for some orders where there are few absorption lines towards the edges of the detector, we fitted 2nd order polynomials instead, to avoid massive deviation of the wavelength solution towards the edges. All these fits were manually inspected to ensure a proper wavelength solution was obtained. For some orders the wavelength solutions were poorly constrained due to strong residual science fringes and/or weak telluric absorption features. For these reasons we excluded orders 7-9 and 11 from the rest of the analysis. The wavelength solution and root mean square (RMS) deviation for each spectral order is shown in Fig.~\ref{fig:wavsolution_15oct2016} for the first observing night.
    \item Throughput correction (see Fig.~\ref{fig:postprocessing}, panel 3): throughput variations can be caused by variations in airmass and misalignment of the target on the slit. We correct for these variations by dividing by the mean of the brightest 50 pixels per spectrum following the procedure described in \citet{Brog19}.
    \item Telluric removal (see Fig.~\ref{fig:postprocessing}, panel 4-5): quasi-stationary trends mostly due to the telluric lines are still dominating the spectra. To remove these, Principal Component Analysis (PCA) is used to decompose the spectral time series. In our implementation, we perform a Singular Value Decomposition \citep[similar to][]{deK13, Line2021}. Visual inspection of other ARIES observations from other systems indicates that using seven Principal Components works well to balance telluric removal and retrieve an injected exoplanet signal at a high S/N (as described in Section~\ref{s:sn-method}), so we adopt this also for WASP-33 b. However, we emphasize that we keep the number PCA iterations fixed for all spectral orders and all observing nights to avoid optimisation of order/night specific systematic effects and noise, which can produce artificially large S/N (e.g. \citealt{Cabot2020,Spring2022}).
    Remaining low-frequency residuals persist due to several effects: (1) throughput variations due to small offsets of the target position on the spectrographs slit entrance; (2) residual airmass variations; (3) echelle traces drift on the detector due to the changing gravity vector during the observations, causing instrumental flexure; and (4) uncorrected science frame fringes. We apply a 6th order Butterworth high-pass filter with a cutoff frequency of $0.02 \ \rm{pixel}^{-1}$ to remove most of these trends.
    
    \item Masking (see Fig.~\ref{fig:postprocessing}, panel 6): remaining bad columns in the residual time series are masked. This is done by identifying $> 3 \sigma$ outliers in the residual matrix and flagging columns with more than five outliers. We combine flagged columns if there are more than two columns within a five pixel-wide sliding window. Additionally, the columns in a 50 pixel window from the edge of the detector are masked by default.
\end{enumerate}
The final result of the post-processing pipeline is shown in panel 6 of Fig.~\ref{fig:postprocessing}.

\subsection{Observed instrument performance}
\label{sec:obs_inst_perf}
\begin{figure}
    \centering
    \includegraphics[width=\columnwidth]{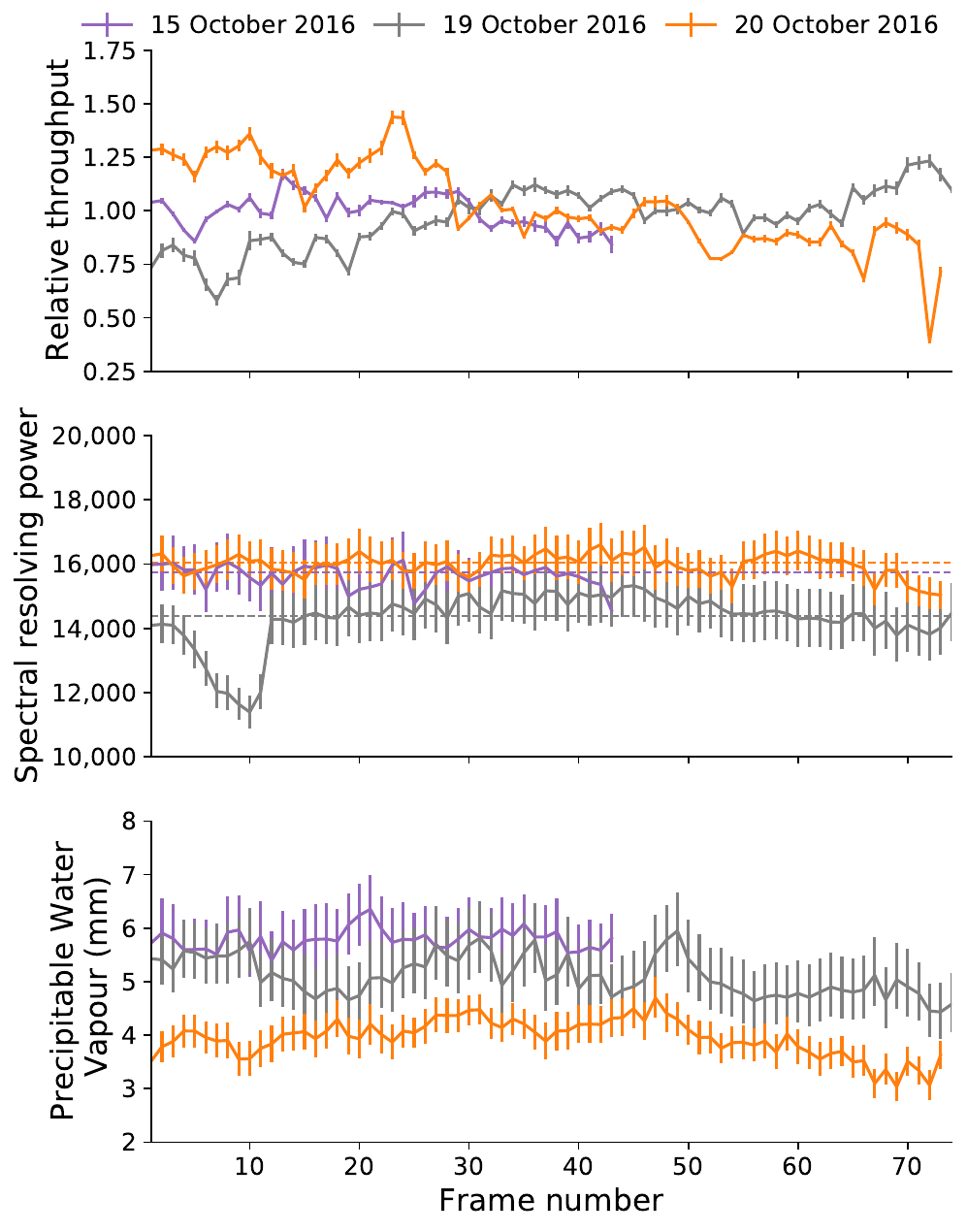}
    \caption{Relative throughput, spectral resolving power and PWV for each observing night, as measured from order 23 which is centered around the CO 2.3 $\micron$ spectral feature. The dashed lines indicate the median resolving power used to convolve the spectral templates for each night. The overall consistently reduced resolution of ARIES is hypothesised as sub-optimal focusing of the spectrograph.}
    \label{fig:measured_IP}
\end{figure}

As this is the first time ARIES/MMT is used to characterize an exoplanet atmosphere, we compare the observed performance to the theoretical instrumental performance. We measured the instrumental throughput, resolving power and Precipitable Water Vapour (PWV) directly from the normalised spectral time series. The instrumental throughput is measured directly from the total S/N for each spectra. To measure the observed resolving power and PWV we largely follow procedure described in \cite{Chiavassa2019}, where we cross-correlate a telluric model with the normalised spectral time series and evaluate the log-Likelihood defined by \cite{Zucker2003}. This method is sensitive to continuum variations and first we correct for them the following way: (1) a median filter removes any possible outliers (2) each spectrum is binned and the maximum of each bin is calculated (3) we fit a low 3rd-order polynomial to these local maxima and (4) we divide out this polynomial fit from each spectra. The {\sc Telfit}\footnote{https://github.com/kgullikson88/Telluric-Fitter} code is used to compute telluric spectra \citep{Gullikson2014}. We fix the airmass to the logged airmass for each frame which we obtain from the raw FITS-header information, instead of keeping it a free parameter as done by \cite{Chiavassa2019}. The Markov Chain Monte Carlo (MCMC) {\sc PyMultiNest}\footnote{\url{https://johannesbuchner.github.io/PyMultiNest/index.html\#citing-pymultinest}} \citep{Buchner2014} package is used to constrain PWV and R.
The measured relative throughput, instrumental resolving power and PWV are shown in Fig.~\ref{fig:measured_IP}. A key results is that the measured spectral resolving power is $R = 14, 000 - 16,000$, notably lower than the nominal specification for ARIES of $R = 30,000$. We hypothesize this large offset may be due to sub-optimal focus of the spectrograph resulting in consistently lower observed spectral resolution. The other fluctuations of spectral resolving power in the course of each night could be due to a combination of target misalignment on the slit entrance, instrumental instability and occasional opening of the AO loop. This analysis also highlights the importance of measuring the spectral resolving power from the observations when using HRCCS, as our cross-correlation templates will need to be convolved to the appropriate spectral resolution later on.

\section{Atmospheric models}\label{sec:modeling}
As we use the HRCCS technique to characterise the atmosphere of WASP-33 b, we require forward models for the cross-correlation templates i.e. synthetic spectra of WASP-33 b, which we discuss in this section. 
%
We use several atmospheric modeling codes in this work to address the structure, chemistry, and dynamics of the atmosphere of WASP-33 b. To identify absorption and emission features initially, we used PHOENIX atmospheric models. PHOENIX is a general purpose atmosphere code, well-tested on objects from cool planets to hot stars \citep[e.g.][]{Hauschildt1997,Hauschildt1999,Barman2001,Barman2011,Allard2011,Lothringer2018,Lothringer2019}. We used both self-consistent atmosphere models in Section~\ref{sec:self_consistent_models} in 
thermodynamic equilibrium, and a grid of models where the temperature structure and/or the molecular abundances have been manually varied in Section~\ref{sec:phoenix_mod}. We then further use the Monte Carlo Radiative Transfer code gCMCRT \citep{Lee2019,Lee2021} to model 3D effects and orbital phase-dependent variations of the observed spectra in Section~\ref{sec:gcm_modeling}.

\subsection{Self-consistent PHOENIX models} \label{sec:self_consistent_models}
\begin{figure*}
    \centering
    \includegraphics[width=\textwidth]{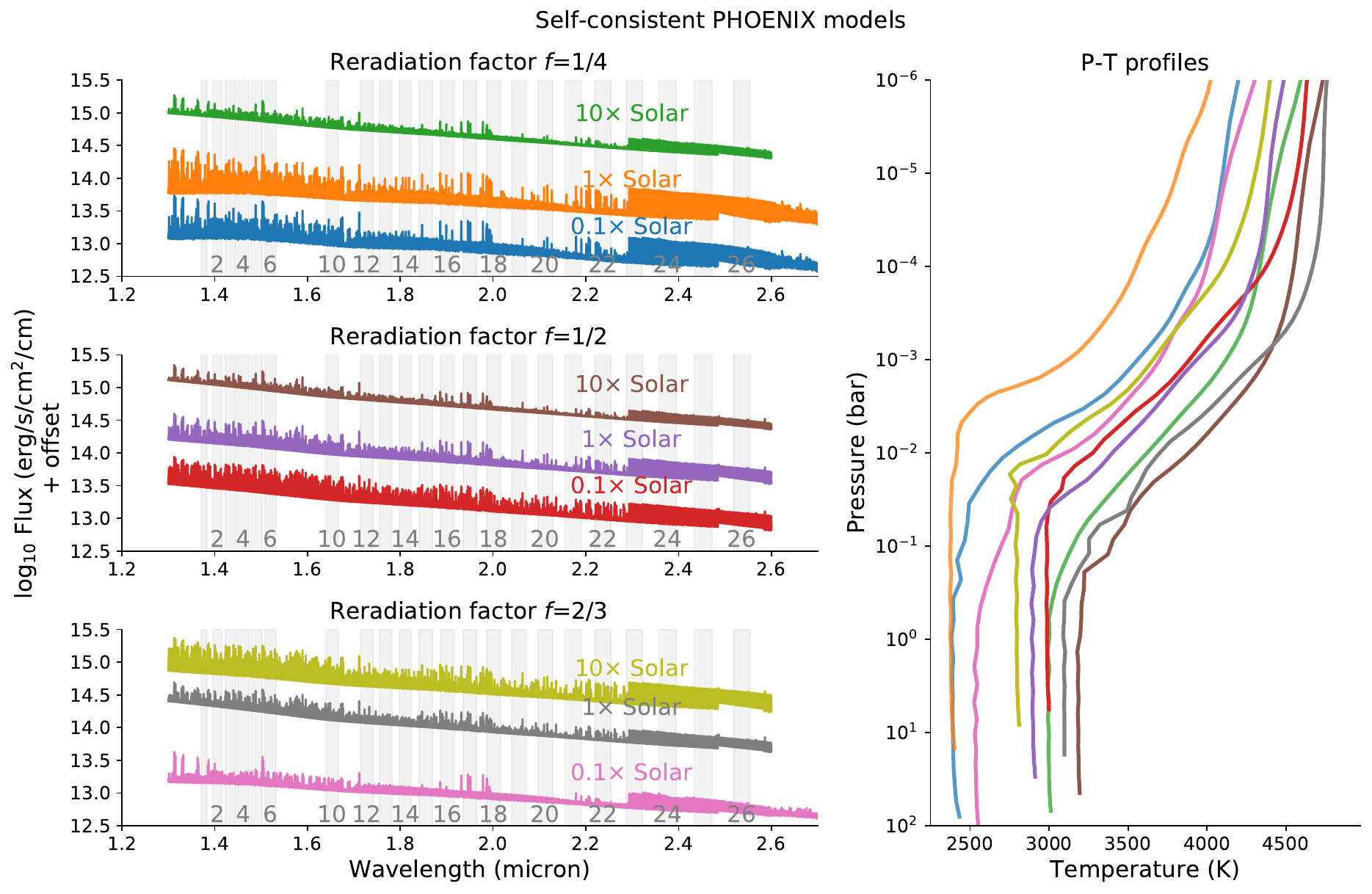}
    \caption{WASP-33 b 1D PHOENIX self-consistent synthetic emission spectra. \textit{Left panels}: PHOENIX self-consistent spectra as a function of the metallicity and re-radiation factor $f$:  $f = 1/4$ for full heat-redistribution, $f = 1/2$ for day-side-only heat-redistribution and $f = 2/3$ for instantaneous re-radiation. Different metallicity models have been offset by 0.75 along the y-axis with respect to the 0.1$\times$Solar model. The ARIES spectral order wavelength ranges included are indicated by the gray bars. \textit{Right panel:} corresponding PHOENIX self-consistent P-T profiles.}
    \label{fig:phoenix_models_overview}
\end{figure*}

The self-consistent models used here are similar to the extremely irradiated hot Jupiter models presented in \citet{Lothringer2018}. The self-consistent atmosphere models solve the atmosphere structure and composition iteratively, calculating the spectrum at each iteration and comparing it to radiative equilibrium. For our models the radiative-convective boundary is usually below the highest pressure we model and no convective adjustments were made in order to increase speed of convergence. Temperatures are then modified to approach radiative equilibrium after which the chemistry is calculated based on chemical equilibrium and the spectrum is re-calculated. This process is repeated until the temperature corrections are small, indicating that the atmosphere is in radiative and chemical equilibrium.
Each spectrum is calculated from 10 to $10^7$ nm using planetary parameters by \citet{Haynes2015} \citep[as summarised in Table~1 of][]{Lothringer2018} on a grid with varying atmospheric metallicity and a heat re-radiation efficiency $f$ \citep[where $f$ is the parameter as defined in][]{Madhusudhan2009}. For the metallicity, we explore a grid of models at 0.1$\times$, 1$\times$ and 10$\times$ Solar metallicity. For the heat re-radiation factors $f$ we explore the values: $f = 1/4$ for day-side and night-side heat redistribution, $f=1/2$ for day-side-only heat redistribution and $f=2/3$ for instantaneous heat re-radiation.
These self-consistent models include \citep[see][]{Lothringer2018}: bound-free opacity from H, H$^-$, He, C, N, O, Na, Mg, Al, Si, S, Ca, and Fe, free-free opacity from H, Mg, and Si, and scattering from e$_-$, H, He, H$_2$. They also include opacities for TiO and VO, but as explained in \citet{Lothringer2018}, TiO and VO are not the sole cause of temperature inversions in extremely irradiated hot Jupiters. Temperature increase at these pressures is likely due to the absorption of short-wavelength radiation by absorbers like atomic Fe and the lack of coolants such as $\rm{H}_2\rm{O}$. Such hot temperatures result in the thermal dissociation of most molecules in the atmosphere including H$_2$ and H$_2$O (see Fig~\ref{fig:phoenix_mixing_ratios}). The abundance of CO, which exhibits the strongest molecular bond, is also reduced, but to a lesser extent compared to other potential molecular opacity sources. The resulting spectra, shown in Fig~\ref{fig:phoenix_models_overview}, predicts CO emission lines with signatures of H$_2$O and other molecules muted due to thermal dissociation combined with the isothermal lower atmosphere.
\subsection{PHOENIX models with Modified Structure and/or Abundances} 
\label{sec:phoenix_mod}
In addition to the self-consistent modeling, we ran a grid of models with varying temperature structures and atmospheric metallicity to further explore the parameter space. Again, we explore metallicities of 0.1$\times$, 1$\times$, 10$\times$ and 100$\times$ Solar. For the P-T profiles, we use the parameterisation from \citet{Madhusudhan2009}, but modified to have an isothermal upper atmosphere. These simple structures used five parameters to describe the temperature throughout the atmosphere: $P_{1}$, $P_{2}$, $T_{1}$, $T_{2}$ and $\alpha_2$. $P_{1}$ describes the pressure at the tropopause, above which the temperature was set to $T _{1}$. The temperature then varies until it reaches temperature $T_{2}$ at pressure $P_2$, as set by the gradient of the inversion $\alpha_2$. 
We explored a wide range of P-T profiles with upper atmosphere temperatures from 2000 to about 10,100 K, shown in Fig.~\ref{fig:phoenix_modified_structures}. In total we ran 576 forward models.
We also generated a small subset of these models to explore how removing a single opacity source affects the significance of the planet detection i.e we re-run the best matching model without H$_2$O, without CO, and without OH. We do this in order to identify potential absorbers or emitters in the atmosphere, while keeping the P-T profile fixed. Line lists for the investigated opacity sources were taken from: CO \citep{Goorvitch1994}, OH \citep{Barber2006} and H2O \citep{Roth09}.

\subsection{Global circulation model (GCM) and post-processing modelling}
\label{sec:gcm_modeling}

For the 3D GCM, we use output from a SPARC/MITgcm \citep{Showman2009} of WASP-33 b with similar set-up to the simulations performed in the \citet{Parmentier2018} study for UHJs.
The GCM output is then post-processed using the gCMCRT 3D RT code \citep{Lee2021} to provide high-resolution emission spectra that account for variation in the spectra throughout the orbital phases of the observation. We use line-list data from various sources, namely: OH \citep{Hargreaves2019}, H$_{2}$O \citep{Polyansky2018}, CH$_{4}$ \citep{Hargreaves2020}, CO \citep{Li2015}, CO$_{2}$ \citep{Yurchenko2020}, NH$_{3}$ \citep{Coles2019}, HCN \citep{Barber2014}. Cross-sections were calculated using the HELIOS-K opacity code \citep{Grimm2021} and interpolated between 1.27-2.66 $\mu$m at a resolution of R = 100,000.

\begin{figure*}
    \centering
    \includegraphics[width=\textwidth]{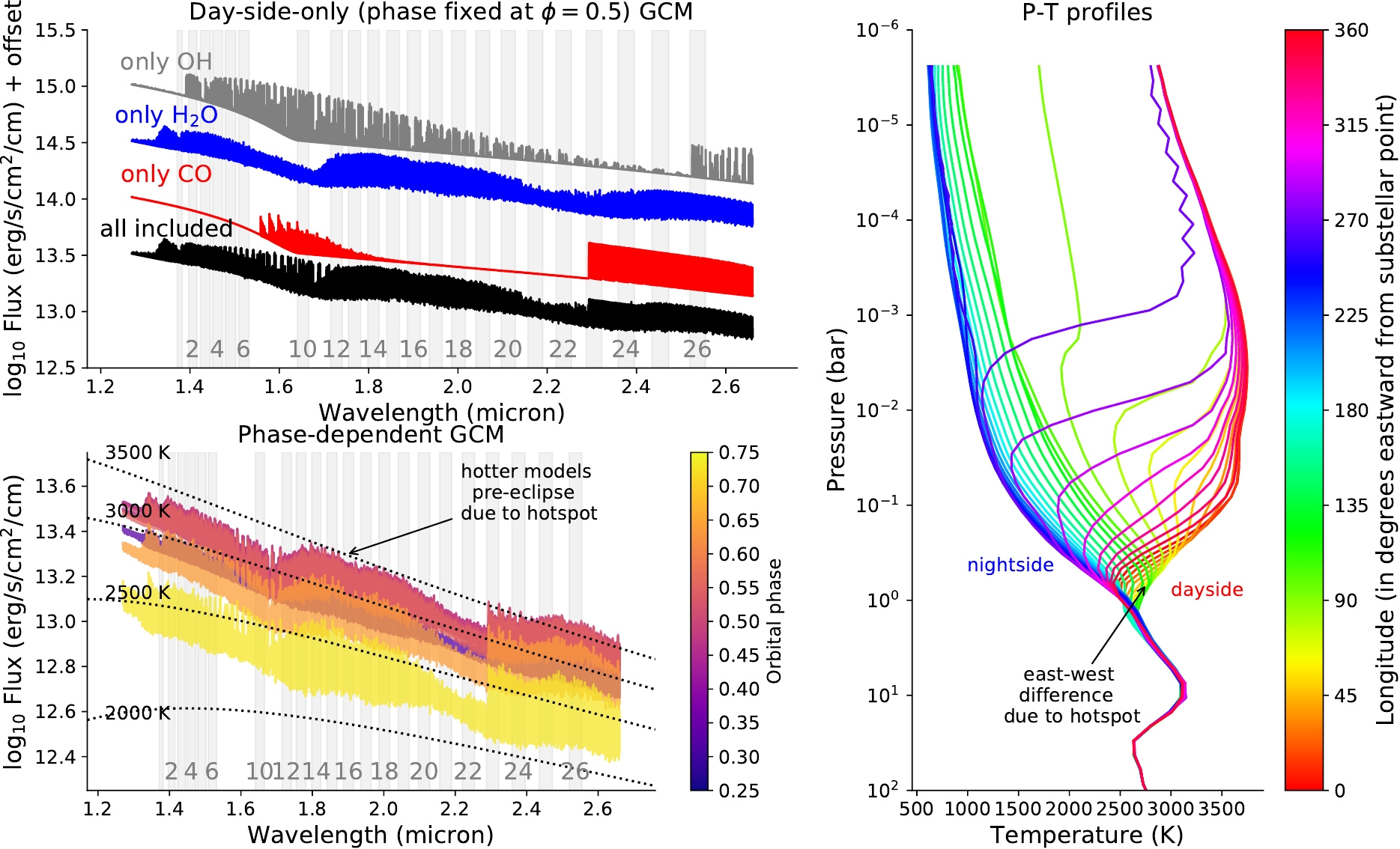}
    \caption{WASP-33 b GCM synthetic emission spectra. \textit{Top-left panel}: Day-side-only (i.e. fixed at phase $\phi=0.5$) GCM spectra with all opacity sources included or only specific molecules included. The different models have been offset by 0.5 along the y-axis with respect to the spectrum at the bottom. The ARIES spectral order window wavelength ranges included are indicated by the gray bars. \textit{Bottom-left panel}: phase-dependent GCM spectra at different orbital phases. Planck black body curves are plotted at a range of temperatures. The GCM predicts hotter models pre-eclipse compared to post-eclipse due to an hotspot approximately $+8 \degree$ eastward seen from the substellar point. \textit{Right panel:} GCM P-T profiles as a function of longitude, where $0^\circ$ is defined as the substellar point and increases towards the eastward direction as seen from the substellar point.}
    \label{fig:gcm_models_overview}
\end{figure*}

The resulting synthetic high-resolution emission spectra include the Doppler shifting of spectral lines due to winds and planetary rotation towards the line of sight for each phase.
We include all molecular species listed above as well as CIA H${2}$-H-He pair continuum opacities from \citet{Karman2019} and H$^{-}$ from \citet{John1988}.
The resulting phase-dependent GCM spectra and P-T profiles of the gCMCRT are shown in the bottom-left and right panels of Fig.~\ref{fig:gcm_models_overview}. It is computationally expensive and resource intensive to run the spectral calculation, as it uses a Monte-Carlo radiative transfer method on many GPUs. Therefore, we computed synthetic spectra at nine phases uniformly sampled over the phase coverage during our three observing nights. For those spectral templates required at intermediate phases, we linearly interpolate these spectral templates in phase. The GCM predicts an eastward hot spot at $+8 \degree$ east from the substellar point, which results in an asymmetry between P-T profiles with an equal longitudinal distance west or east as seen from the sub-stellar point. As the planet rotates in the same direction as it orbits its host star, we see more of the hot spot region pre-eclipse compared to post-eclipse. This results in phase-dependent variations in both the continuum and relative line strengths (see bottom-left panel of Fig.~\ref{fig:gcm_models_overview}). As noted by \citet{deKok2014}, while the overall flux on cooler night sides may be lower, the relative line-to-continuum strength, which HRCCS is sensitive to, is strong. This suggests then that the phase-dependant change in the relative line strengths from the GCM will also be detectable as WASP-33 b proceeds in it orbit.
To enable comparison and cross-check between the 3D GCM and the 1D PHOENIX model atmospheres, we also created a set of day-side-only GCM spectra i.e. fixed at orbital phase $\phi=0.5$ (see top-left panel of Fig.~\ref{fig:gcm_models_overview}). These templates were created by simulating the emission assuming continuum opacity without the effects of Doppler shifting. We also again explore the impact of any single absorber or emitter on the planet signal by creating day-side-only templates with only a single opacity source included i.e. only CO, only H$_{2}$O and only OH. CO is most prolific in spectral lines in the reddest parts of the ARIES spectra, while OH dominates its shorter wavelengths, and water persists throughout. 

\section{High-resolution Cross Correlation Spectroscopy}
\label{sec:hrccs}
The HRCCS technique has been developed in recent years to enable not simply detection of molecular species, i.e. the S/N-method, but into a framework that enables statistically rigorous assessment of atmospheric models that may match the data e.g. Cross-Correlation-to-log-Likelihood (CC-to-log(L)) mapping \citep[e.g.][]{Brog19,Gibson2021}. Both methods have different uses, and we use both in this work to characterise the composition, structure, and dynamics of the WASP-33 b atmosphere, as described below.

\subsection{S/N method}
\label{s:sn-method}
\begin{figure*}
    \centering
    \includegraphics[width=\textwidth]{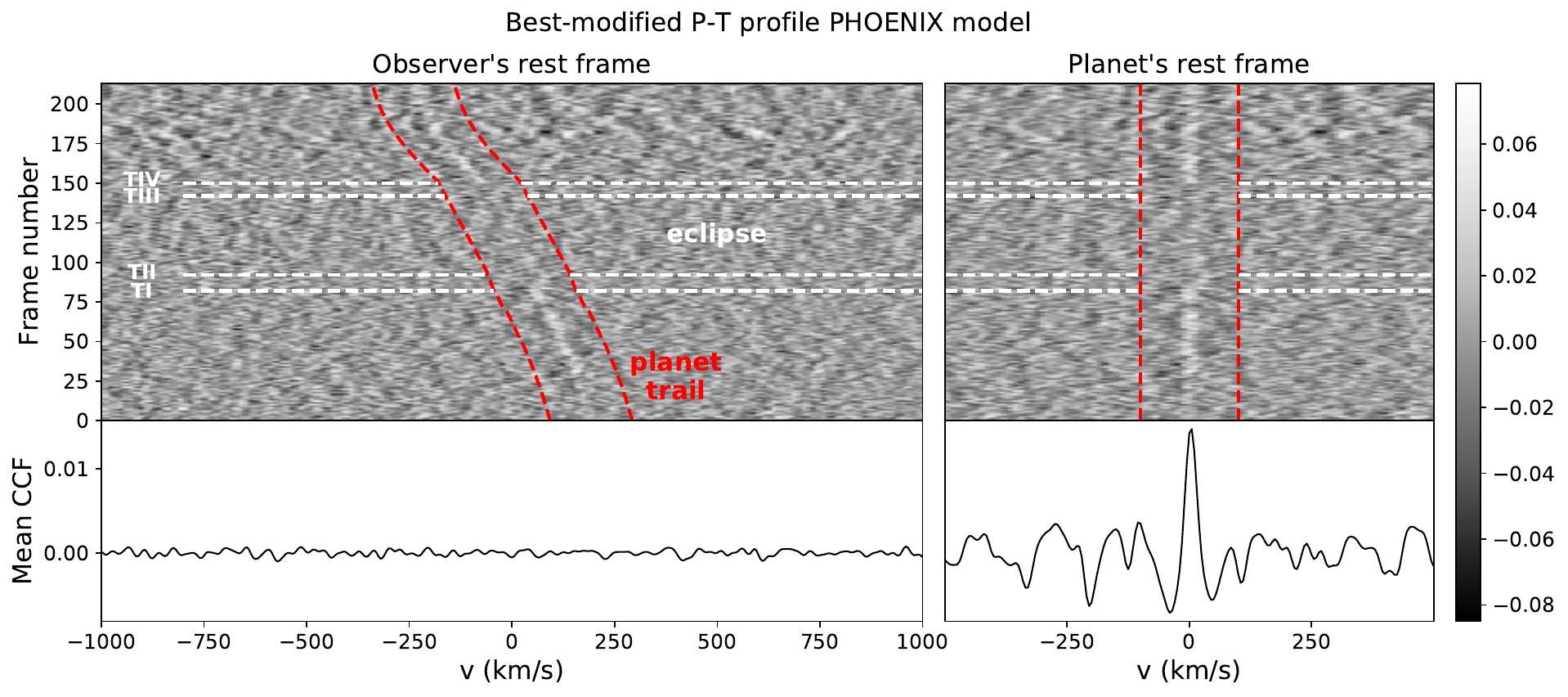}
    \caption{Cross-Correlation Functions for the best-modified P-T profile detected at a S/N of 7.9$\sigma$. \textit{Top-left:} The CCF for each frame including all three observing nights. Frames are sorted by orbital phase but are not equally spaced, and the blue dotted line marks the post-eclipse frame with symmetric phase to the earliest pre-eclipse frame (see Figure~\ref{fig:phase_coverage}). The start/end of ingress/egress are indicated by the white dashed lines labeled TI-TIV for the four contact points. The two red lines are centered around the planet radial velocity trail. \textit{Top-right:} Same as the top-left panel, but shifted to the planet rest frame. \textit{Bottom-left:} the mean CCF along the y-axis. Frames where the planet is not fully visible (in between TI-TIV) were excluded when computing the CCF to prevent dilution of the planet signal. \textit{Bottom-right:} same as the bottom-right panel, but shifted to the planet's rest frame.}
    \label{fig:phoenix_cc_trail}
\end{figure*}

The S/N-method cross-correlates each observed spectrum with a spectral template and the S/N is calculated from the maximum and standard deviation of the corresponding Cross-Correlation Function. First the spectral template is scaled to the expected star-planet contrast ratio using
\begin{equation}
    {F_{\rm{model,scaled}} = \frac{F_{\rm{model}}}{B_{\rm{\star}}} \frac{{R_{\rm{p}}}^2}{{R_{\star}}^2}},
    \label{eq:fscaled}
\end{equation}
with planet radius $R_{\rm{p}}$, stellar radius $R_{\rm{star}}$ and assuming a stellar black body flux $B_{\rm{\star}}$ at the effective stellar temperature (using values in Table~\ref{tab:wasp33_system}). A stellar black body is a reasonable assumption for the host star WASP-33 given its hot effective temperature which results in few spectral stellar absorption lines in the wavelength region of interest. Previous works indicate that it is important to match the line shape correctly for HRCCS \citep[e.g.][]{Spring2022}, hence before cross-correlating, we convolve each synthetic spectrum of WASP-33 b to the median observed resolving power for each night (see Fig.~\ref{fig:measured_IP}), using a Gaussian instrumental profile. Since we applied a 6th-order Butterworth high-pass filter at a cutoff-frequency of $0.02 \ \rm{pixel}^{-1}$ to our data, we also apply it to the scaled model template, where we adjust the cutoff-frequency to account for the difference in sampling rate between the model and data wavelength grid. We calculate the correlation coefficient as defined by \citet{Brog19} for each observed residual for each spectral order for all observing nights with the scaled template. We explore a radial velocity range of [-1000, 1000] km/s in steps of $\Delta v = 5 \ \rm{km/s}$, close to the theoretical instrumental resolution of ARIES. This results in a CCF time series for each spectral order, we will refer to this as the CCF-matrix (see Figure~\ref{fig:phoenix_cc_trail}). To robustly detect if the emission source is indeed the planet, it is common practice to shift each CCF-matrix using a grid of trial velocities into the proposed planet rest frame velocity (see right panel of Figure~\ref{fig:phoenix_cc_trail}):
\begin{equation}
    {v_{\rm{trial}} = v_{\rm{bary}} + v_{\rm{sys}} + K_{\rm{p}} \sin{\phi_{\rm{p}}}},
    \label{eq:vtrial}
\end{equation}
with $v_{\rm{bary}}$ the barycentric velocity, $v_{\rm{sys}}$ the trial system velocity, trial $K_{\rm{p}}$ the semi-major amplitude of the planet's orbital velocity and $\phi_{\rm{p}}$ the planet's orbital phase, where we have assumed a circular Keplerian orbit. The Python package {\sc{Barycorrpy}}\footnote{https://github.com/shbhuk/barycorrpy} \citep{Wright2014,Kanodia2018} is used to calculate $v_{\rm{bary}}$ at the observed times. We explore a $31 \times 31$ grid in $(v_{\rm{sys}}, K_{\rm{p}})$-space around the expected values of ($v_{\rm{sys}}$, $K_{\rm{p}}$) = (-0.3, 230.9) \rm{km/s} as previously found in the detection of OH by \citet{Nugroho2020} within the range of $\pm 100$ km/s for $v_{\rm{sys}}$ and $\pm 150$ km/s for $K_{\rm{p}}$ on a 31x31 grid. For each trial velocity on the grid, we shift the CCF-matrix to $v_{\rm{trial}}$ and calculate the combined CCF by taking the mean along the time axis. The S/N  is determined by calculating the maximum of the combined CCF and dividing it by the standard deviation of the CCF. Those exposures taken in between the first and last contact of the eclipse of WASP-33 b (see Fig.~\ref{fig:phase_coverage}) are excluded from the S/N calculation to avoid dilution of the combined CCF signal.
\begin{figure}
    \centering
    \includegraphics[width=\columnwidth]{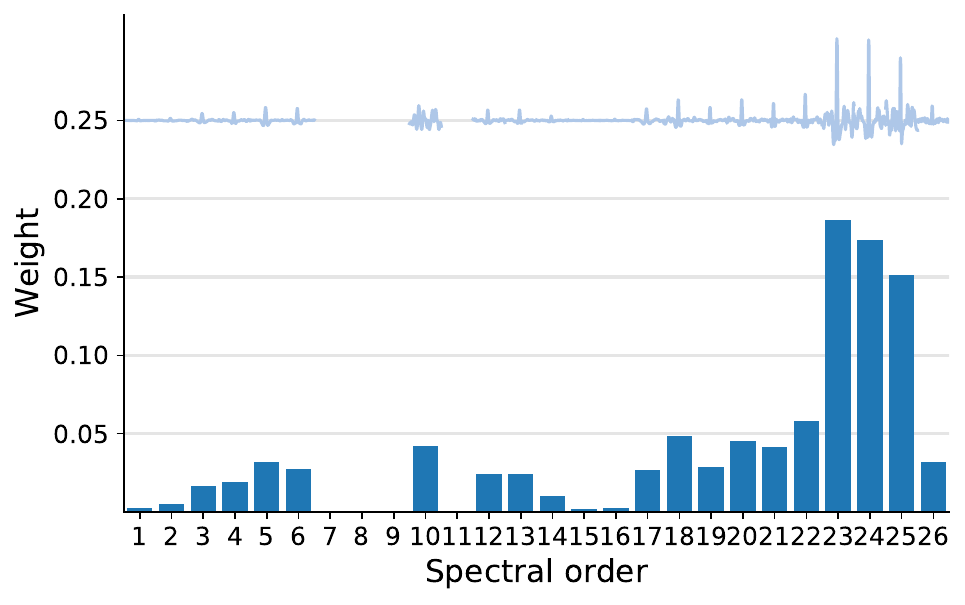}
    \caption{Weights per spectral order after cross-correlation with the best self-consistent PHOENIX spectral template used in our S/N-method adopted. Weights are calculated from the S/N of the difference between the injected and observed CCF, indicated by the miniature CCF above each order. Orders 7-9 and 11 were excluded from the analysis in the wavelength calibration step. Orders 23-25 around the CO 2.3$ \micron$ spectral feature are naturally assigned the highest weights.}
    \label{fig:cc_weights}
\end{figure}
Not all spectral orders are equally sensitive when cross-correlating with the spectral template. Lower sensitivity is expected for the bluer ARIES orders due to poorer data quality due to lower throughput and stronger spectral fringes and weaker spectral features in the models. To account for both, we use an order-specific weighting scheme in the S/N-method to combine the CCF-matrices, based on the data quality and model. To calculate the weights we follow \citet{Spring2022}, where we first inject the model (before PCA cleaning) into the observations using the expected model strength and orbital parameters (using the values reported by \citet{Nugroho2020}). We create a CCF-matrix for each injected order, and then subtract its corresponding observed (no injection) CCF-matrix. For each order we calculate the S/N of this CCF difference. For each order $n$, weights $w_n$ are assigned proportional to this S/N and normalised such that $\sum^{N}_{n=1} w_n = 1$, where $N$ is the total number of spectral orders included. An example of the weighting scheme is shown in Fig.~\ref{fig:cc_weights}. As expected, using the self-consistent PHOENIX at $\times$1 Solar metallicity with a re-radiation factor of $f=1/2$ as an example, the reddest orders around the $\sim$2.3 micron densely-packed CO spectral feature region contribute the highest weights due to the numerous strong spectral lines and higher S/N of the observed spectra.
\subsection{CC-to-log(L) mapping and model scaling parameter}
\label{s:cctologl}
While the S/N-method is widely used in literature and thus a convenient way to compare the significance of our detection with previous works, it is inherently normalised by definition and therefore insensitive to scaling of the model or observed spectra. This is why there was a need to implement a weighting scheme to combine multiple spectral orders in the S/N-method. The CC-to-log(L) Bayesian Framework described by \citet{Brog19} however does not suffer from this. CC-to-log(L) mapping maximizes the log-likelihood as:
\begin{equation}
    {\log{L} = -\frac{N}{2}\log{\big( s_g^2 - 2 R + s_f^2} \big)}
    \label{eq:logL_brogiline}
\end{equation}
with $s_{\rm{g}}$ and $s_{\rm{f}}$ the variance of model and data, respectively, $R$ the cross-covariance and $N$ the number of spectral channels. When combining multiple nights and spectral orders, we can compute the total sum of their log-likelihood values. There is thus no need to weight each spectral order as this is accounted for in the CC-to-log(L) mapping \citep[][]{Brog19}. Moreover, CC-to-log(L) mapping gives us a framework to extract uncertainties on the measured and derived quantities from the modelling.

Our implementation uses {\sc Pymultinest}\footnote{https://johannesbuchner.github.io/PyMultiNest/} \citep{Buch14}, a generic Python package connected to {\sc MultiNest}, a Bayesian inference tool\footnote{https://github.com/JohannesBuchner/MultiNest} \citep{Feroz08, Feroz09, Feroz19}, for all our parameter estimations. Importantly, following \citet{Brog19}, we also introduce one additional free parameter, the scaling factor $a$, which allows us to compensate for any potential unknown scaling of the model template, but ideally should be retrieved at $a=1$. We multiply the template by $a$ after applying the high-pass filter. The multidimensional parameter space is explored using the priors in Table~\ref{tab:bl19priors}.
\begin{table}
    \centering
    \begin{tabular}{lcc}
        \hline
         Parameter&Symbol&Prior  \\
         \hline\hline
         System velocity offset&$\Delta v_{\rm{sys}}$&Uniform(-50,50) km/s \\
         Orbital velocity offset&$\Delta K_{\rm{p}}$&Uniform(-50,50) km/s \\
         Scaling parameter&$a$&LogUniform(-2,2) \\
         \hline
    \end{tabular}
    \caption{Priors used in CC-to-log(L) framework. Velocity offsets are given relative to the previous WASP-33 b velocity values of ($v_{\rm{sys}}$, $K_{\rm{p}}$) = (-0.3, 230.9) km/s as reported by \citet{Nugroho2021}.}
    \label{tab:bl19priors}
\end{table}
To statistically compare models drawn from the same underlying parameter distribution amongst each other we apply Wilks' Theorem \citep{Wilks1938}. In practical terms, this theorem states that the difference in $\log(L)$-values, also known as the Bayes factor, between two models with $n$ free parameters follows a $\chi^2$-distribution with $n$ degrees of freedom \citep[for a concise statistical description see][]{Pino2020}. The confidence interval p-value can now be calculated from the corresponding $\chi^2$-distribution and converted to a familiar $\sigma$-value. This way, by its construction, we can only compare models relative to the best trial model.

%
\section{Results}
\label{sec:results}
We now present the results using both the S/N-method and the CC-to-log(L) mapping for our four different modelling suites described in Section~\ref{sec:modeling}.  We detect the atmosphere of WASP-33 b and measure its properties to varying significance and confidence for the best matching model in each modelling suite. The key results are summarised in Table~\ref{tab:snr_results_overview}. The best-matching model from each suite results in the detection the Doppler-shifting spectrum of the planet, as demonstrated by the radial velocity trail in the CCF-matrices shown in Figures~\ref{fig:phoenix_cc_trail} and~\ref{fig:gcm_trail_plots}. Notably, the trail disappears at the calculated start time of secondary eclipse ingress and reappears at the end of its egress, confirming that the detected signal is associated with the planet's spectrum as it is obscured by the star during secondary eclipse.
\begin{table*}
    \centering
    \begin{tabular}{lcc}
        \hline
         Model&Highest S/N&$\Delta \log(a)$=$\log(a_{\rm{post-eclipse}})$ - $\log(a_{\rm{pre-eclipse}})$\\
         &&(symmetric phase coverage)\\
         \hline\hline
         Modified P-T profile PHOENIX models (1D) & 7.9 &$0.21^{+0.03}_{-0.03}$\\
         Self-consistent PHOENIX models (1D) & 7.6 & $0.24^{+0.03}_{-0.03}$\\
         Day-side-only (fixed orbital phase $\phi=0.5$) GCM (3D) & 7.1 &$0.27^{+0.03}_{-0.04}$\\
         Phase-dependent GCM (3D) & 7.1 &$0.13^{+0.03}_{-0.03}$\\
         \hline
    \end{tabular}
    \caption{Overview of the highest S/N-ratio of the sets of models explored in this work and their corresponding scaling parameter $a$ from the CC-to-log(L) mapping. The final column shows $\Delta\log(a)$ i.e. the difference between the mean of the pre-eclipse (Nights 1 \& 3) scaling parameter and that for the post-eclipse data subset for which we have symmetric phase coverage. The $\log(a)$ scaling parameter has a value of zero when the model is a good description of the data. The phase-dependent GCM gives the smallest discrepancy in $\log(a)$, indicating that it is accounting for the phase-dependence of the scaling parameter. }
    \label{tab:snr_results_overview}
\end{table*}

\subsection{Thermal structure and composition}\label{sec:thermal_comp}
\begin{figure*}
    \centering
    \includegraphics[width=\textwidth]{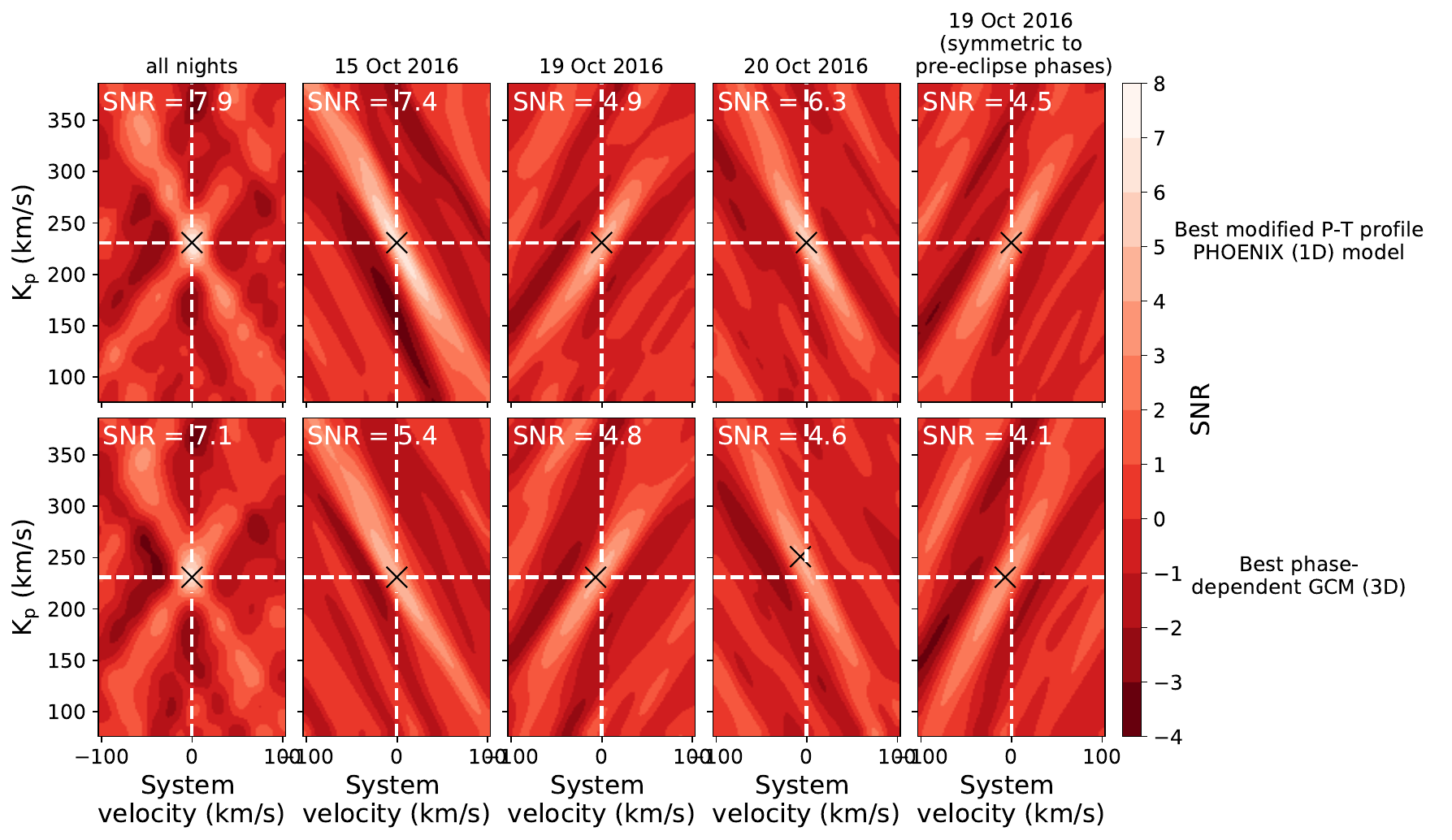}
    \caption{S/N as a function of the system velocity $v_{\rm{sys}}$ and orbital velocity $K_{\rm{p}}$. The gray plus symbol marks the maximal S/N location. The dashed white grid indicates the expected location at ($v_{\rm{sys}}$, $K_{\rm{p}}$) = (-0.3, 230.9) km/s based on the OH detection by \citet{Nugroho2021}. Our detection is in agreement with their ($v_{\rm{sys}}$, $K_{\rm{p}}$)-values for all nights combined. \textit{Top row:} the highest S/N-ratio modified P-T profile PHOENIX model detected at 7.9$\sigma$.
    \textit{Bottom row:} the phase-dependent GCM spectral template detected at 7.1$\sigma$.}
    \label{fig:snr_main}
\end{figure*}

The model that results in the highest S/N detection is a modified PHOENIX model. The highest S/N model includes all opacity sources and has the following P-T profile parameterization: $T_{1} = 3651 \ \textrm{K}$, $T_{3} = 2000 \ \textrm{K}$, $P_1 = 10^{-3} \ \textrm{bar}$, $P_{3} = 10^{-6} \ \textrm{bar}$, $\alpha_{2} = 0.17$ and a metallicity of $10 \times$ Solar. As shown in the top row of Fig.~\ref{fig:snr_main} this model gives $S/N=7.9\sigma$ when all nights are combined, and is higher during pre-eclipse orbital phases than post-eclipse.
In the calculation of the S/N-ratio (i.e. CCF peak / CCF standard deviation), we assume the distribution of the cross correlation values is Gaussian and not impacted significantly by any residual correlated noise. To assess this, we plot the distribution of the cross-correlation values in-trail (three most central columns) and out-of-trail (outside of the ten most central columns). The result, based on the best matching model described above in this section, is shown in Fig.~\ref{fig:wasp33_phoenix_best_modified_trails_hist} for the highest S/N-ratio modified P-T profile PHOENIX spectral template. To quantify whether the samples are normally distributed, we performed a Shapiro Wilks Test on the in-trail and out-of-trail samples (that is after aligning to the planet's rest frame). Both samples pass the test with p-values $>0.05$ ($p_{\rm{in}}=0.75$ and $p_{\rm{out}}=0.09$). Therefore, the distribution of the cross-correlation values follows a normal distribution, and we conclude the effect of residual correlated noise on the S/N-measurement is negligible. 
The correspondingly similar results for the PHOENIX self-consistent models are shown in Fig.~\ref{fig:snr_appendix}. Although the S/N of the detection varies with phase, this alone is not sufficient evidence to interpret it as due to a planetary feature e.g. an offset hot spot, day-to-night variations in brightness, abundances, or thermal structure. However, we discuss in Section~\ref{sec:eastward_hotspot} the additional evidence supplied by the CC-to-log(L) mapping scaling parameter that could support such interpretations.

\begin{figure*}
    \centering
    \includegraphics[width=0.49\textwidth]{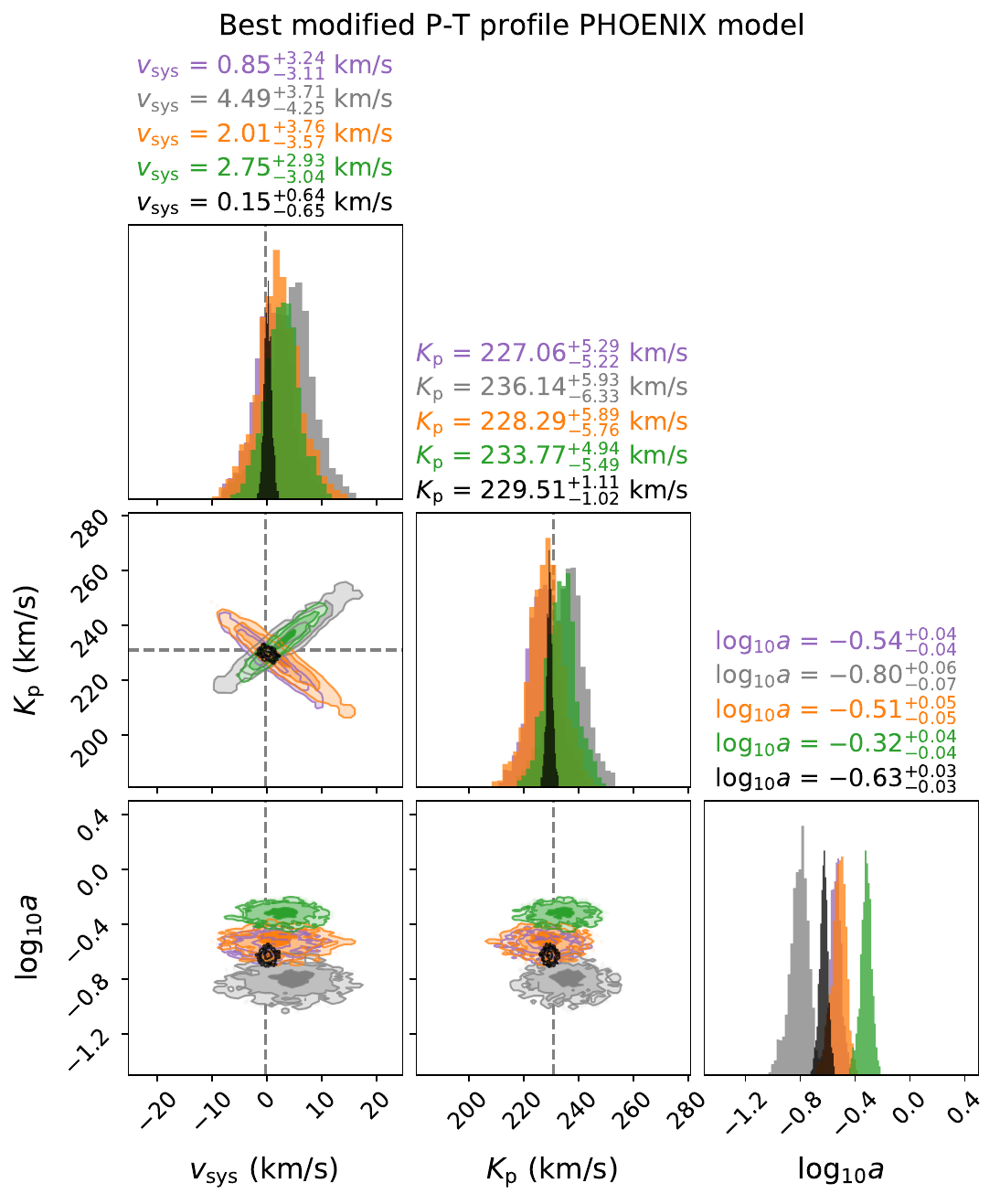}
    \includegraphics[width=0.49\textwidth]{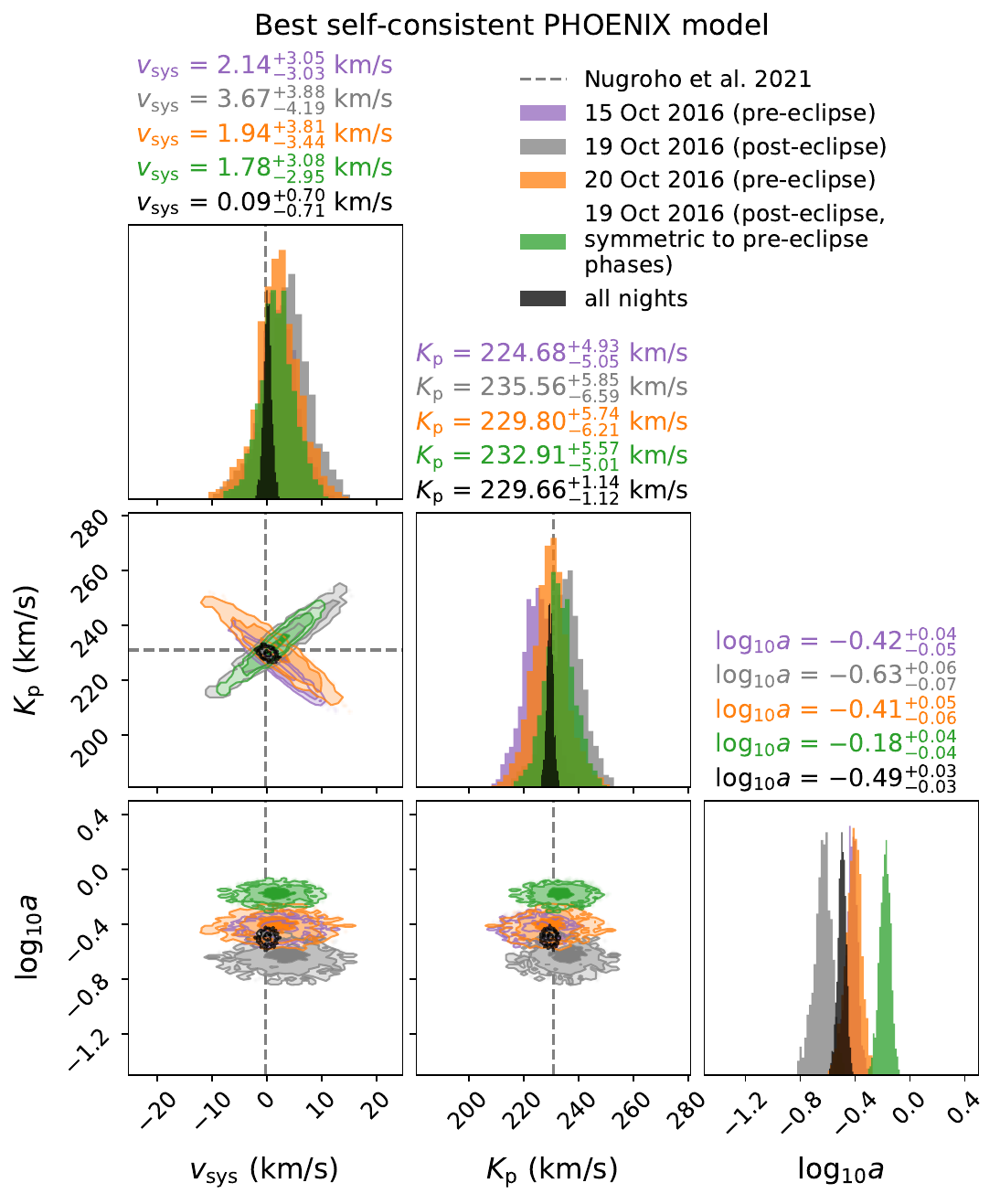}
    \includegraphics[width=0.49\textwidth]{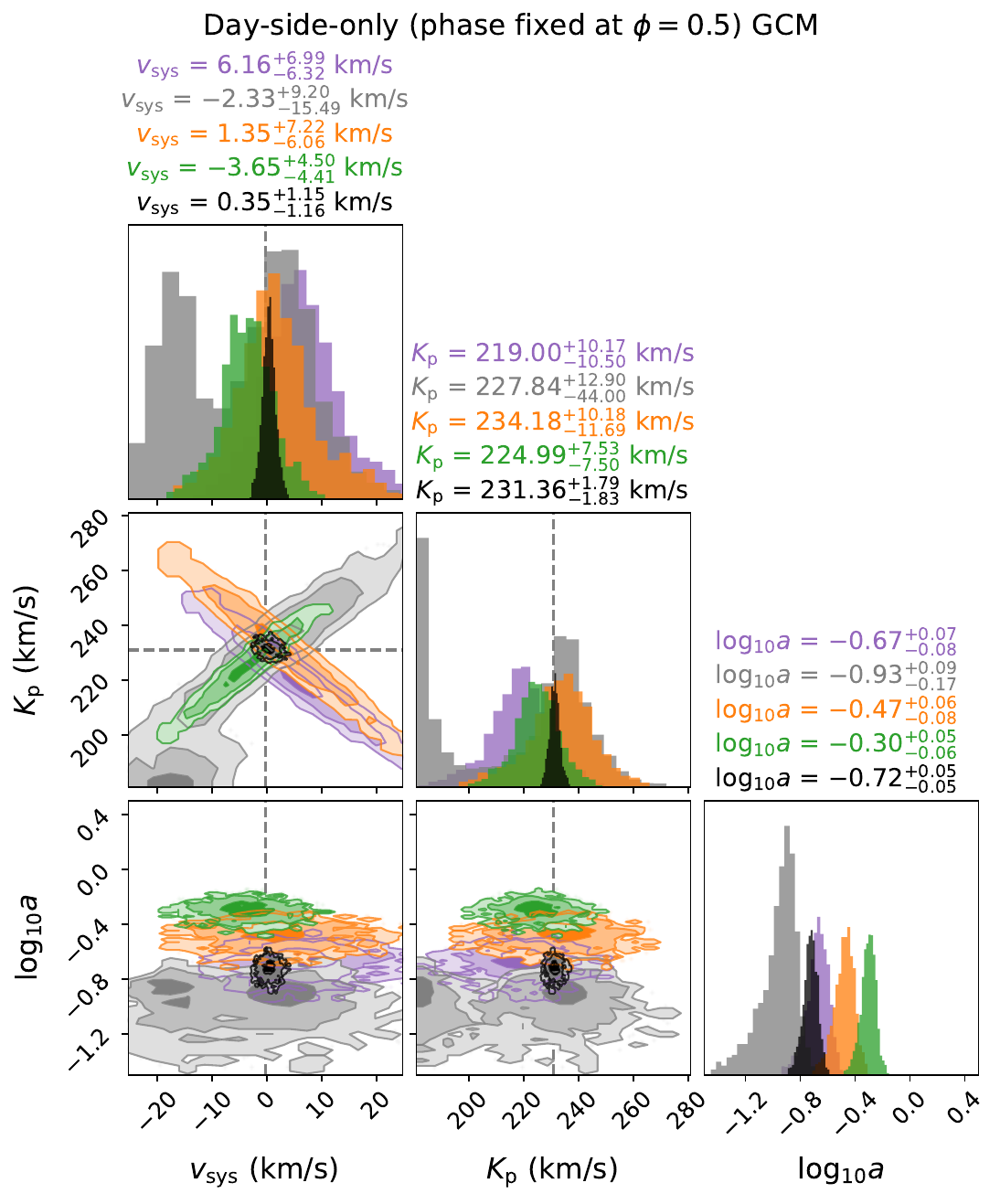}
    \includegraphics[width=0.49\textwidth]{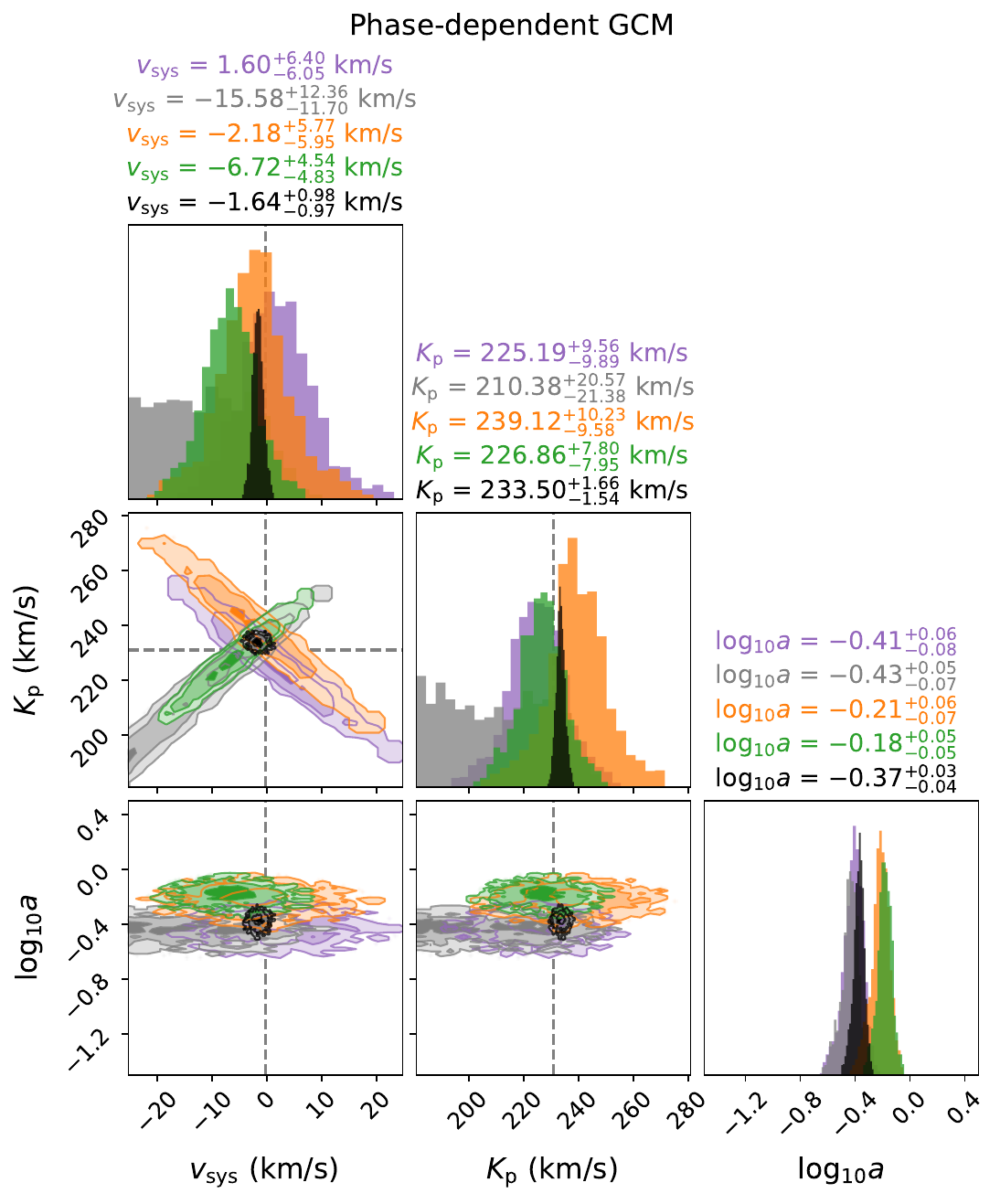}
    \caption{Marginalized distributions from the CC-to-log(L) framework based on  \citet{Brog19}. Results are shown for the best-matching modified P-T profile PHOENIX model (top-left panel), the best-matching self-consistent PHOENIX model (top-right panel), the day-side only GCM (phase fixed at $\phi=0.5$; bottom-left panel) and the phase-dependent GCM (bottom-right panel). The contours indicate the 1$\sigma$, 2$\sigma$ and 3$\sigma$ confidence intervals. All frames with phases during the eclipse have been excluded from this analysis to prevent dilution of the planetary signal. Results are shown for individual observing nights, all nights combined, and for the subset of post-eclipse phases that are symmetrically matched with the pre-eclipse phases (see Fig.~\ref{fig:phase_coverage}). The dashed black grid indicates the expected location at ($v_{\rm{sys}}$, $K_{\rm{p}}$) = (-0.3, 230.9) km/s based on the OH detection by \citet{Nugroho2021}.}
    \label{fig:corner_all}
\end{figure*}

The top-left panel of Fig.~\ref{fig:corner_all} shows the results from the log(L)-to-CC mapping for the best-matching modified PHOENIX model, giving the posterior distributions for the systemic velocity, orbital velocity, and the scaling parameter. The planet is detected at $v_{\rm{sys}} = 0.15^{+0.64}_{-0.65} \ \rm{km/s}$ and $K_{\rm{p}} = 229.53^{+1.11}_{-1.02} \ \rm{km/s}$ with this highest S/N spectral template, in agreement with previous literature e.g. ($v_{\rm{sys}}$, $K_{\rm{p}}$) = (-0.3, 230.9) km/s \citep[][]{Nugroho2021}. The scale factor posterior distribution, which would have $\log(a)=0$ for a perfectly matched model, is systematically offset to smaller values, indicating that the model does not fully encapsulate all of the physics and chemistry to describe the atmosphere. The same thing is seen in the top right panel of Fig.~\ref{fig:corner_all} for the best-matching self-consistent PHOENIX model. However, all nights are in agreement with their ($v_{\rm{sys}}$, $K_{\rm{p}}$)-values within 2$\sigma$. We discuss the variation of $\log(a)$ in later sections.

\begin{figure}
    \centering
    \includegraphics[width=\columnwidth]{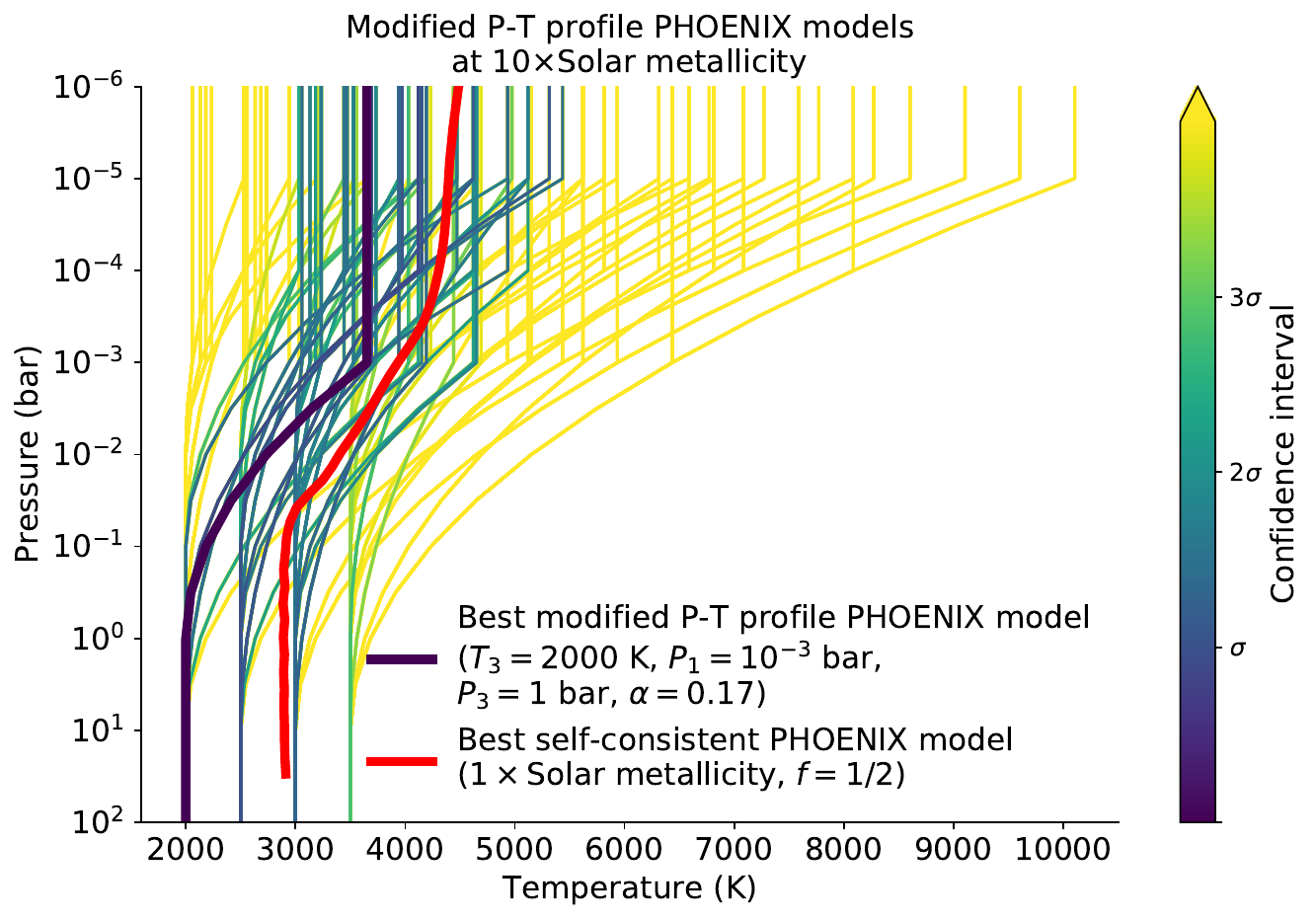}
    \caption{Confidence intervals for the modified P-T profile PHOENIX models at $10 \times$ Solar metallicity. These are for all observing nights combined and were calculated using Wilks' Theorem. The P-T profile of the best modified structure model, the one with the highest log-likelihood, is plotted in dark green and has the following parameterisation: $T_{3} = 2000 \ \textrm{K}$, $P_1 = 10^{-3} \ \textrm{bar}$, $P_{3} = 1 \ \textrm{bar}$, $\alpha_{2} = 0.17$. All confidence intervals here are with respect to this best model. The confidence intervals are converted to their corresponding $\sigma$-value (as described in the last paragraph of Section~\ref{s:cctologl}.) The best self-consistent PHOENIX model's P-T profile, at $1\times$ Solar metalictiy and with a heat reradiation factor $f=1/2$, is plotted in red for comparison.}
    \label{fig:pt_profle_confidence_intervals}
\end{figure}

We compare the best-matching modified P-T profile PHOENIX model with the rest of the models in the suite, as well as that of the best-matching self-consistent PHOENIX model, to give confidence intervals with respect to the alternative P-T profiles, fixed at 10 $\times$ Solar metallicity. This was calculated using Wilks' Theorem (with seven free parameters), and is shown in Fig.~\ref{fig:pt_profle_confidence_intervals}. The best-matching modified P-T profile is slightly cooler than the best-matching self-consistent PHOENIX model, and inverted P-T profiles with an upper and lower atmosphere temperature difference close to the best-matching model are favoured. Both can be understood by the fact that we are sensitive to the the relative line strength with respect to the continuum, which is set by the upper and lower atmosphere temperature contrast (or equivalently the strength of the inversion layer). On the contrary, we are less sensitive to the absolute temperature due to loss of the continuum information in the HRCCS processing, resulting in a range absolute lower and upper atmosphere temperatures inside the calculated 1$\sigma$-confidence interval.

\begin{figure*}
    \centering
    \includegraphics[width=0.75\textwidth]{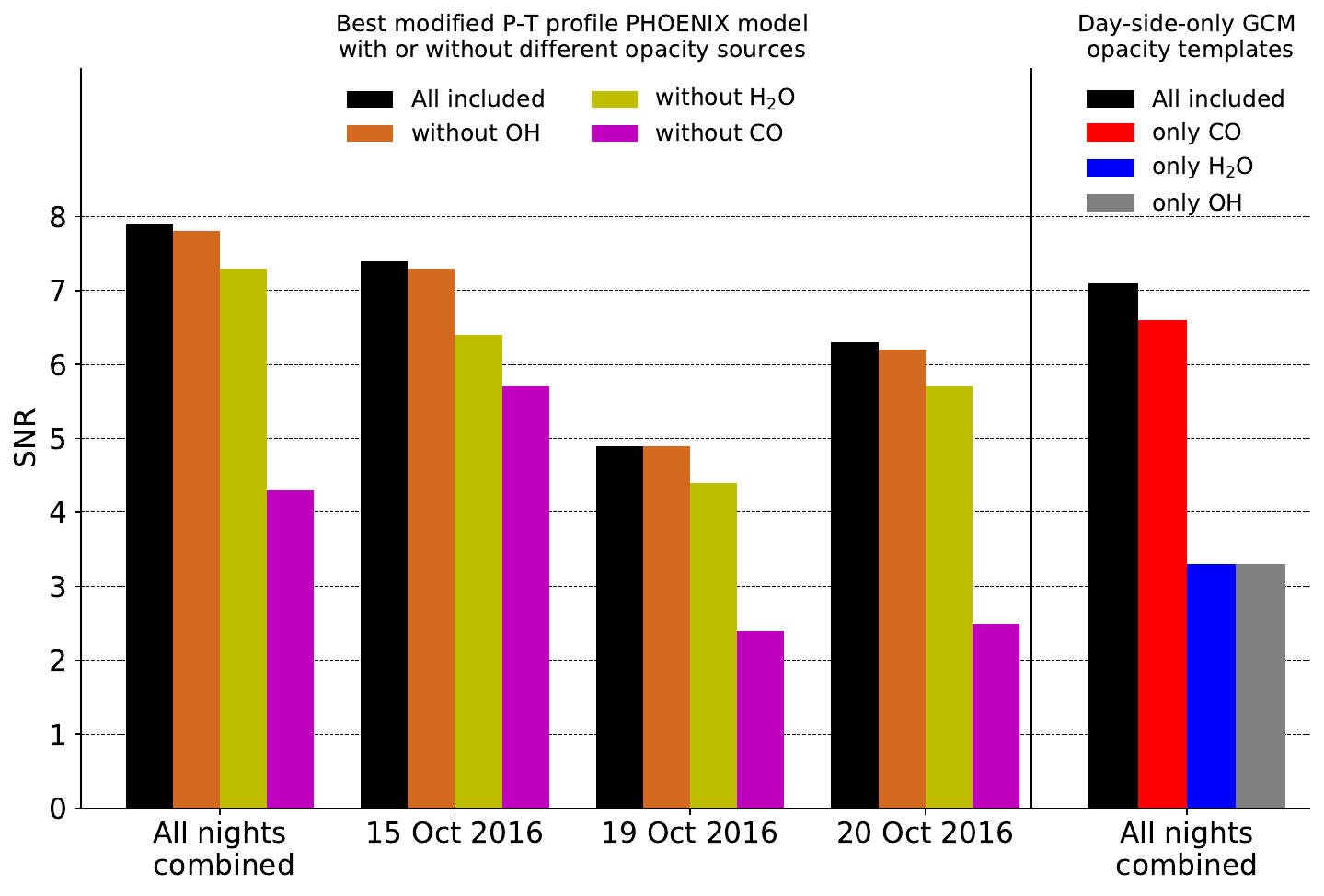}
    \caption{S/N results for the models with different opacity sources included or excluded. \textit{Left panel}: results from the best modified P-T profile PHOENIX models for the cases of all opacity sources included and with a single opacity source excluded, respectively models with OH, $\rm{H}_{2}O$ and CO excluded. All these models fix the P-T profile to the best modified structure PHOENIX model where all opacity sources were included. Only the models without CO result in a significant drop of the S/N when compared to the all included case. \textit{Right panel}: Results from the dayside-only GCM for the cases of all opacity sources included and respectively with only CO, only $\rm{H}_{2}O$ and only OH included. From all opacity sources, only CO remains robustly detected.}
    \label{fig:snr_phoenix_modified_opacities}
\end{figure*}

To determine if any one particular species was contributing the majority of the detected signal, we fix the P-T profile to the best-matching modified P-T profile PHOENIX model (all opacities included), and then run the S/N-analysis for each individual night without CO, $\rm{H}_{2}O$ and OH, respectively. The results are shown in Fig.~\ref{fig:snr_phoenix_modified_opacities}. Exclusion of CO significantly impacts the S/N on every night, and drops from $7.9 \sigma$ to $4.3 \sigma$ for all nights combined. On the other hand, exclusion of OH and $\rm{H}_{2}O$ only marginally reduces the measured S/N, indicating that we are not sensitive to these species at their abundance with this data. However, on the first night of pre-eclipse data, we still detect the planet signal to relatively high S/N, despite the exclusion of CO (the corresponding S/N matrices are shown in Fig.~\ref{fig:snr_noCO}). This suggests that other species in the model are summing to give a detection, but we do not identify these individual species. The data on the third night, which covers the same orbital phase, is of lower quality and may explain why a similar feature is not seen during this night. We also run the same analysis using Wilks' Theorem to compare the models with a single opacity source excluded to the model with all opacity sources included for the best modified P-T profile PHOENIX model (see left-panel of Figure~\ref{fig:opacities_wilks}. We find for all nights combined models without CO are significantly worse ($\sigma=11.1$). Interestingly, there also seems evidence models without H$_{2}$O ($\sigma=5.4$) and marginal evidence OH ($\sigma=3.4$) are worse which we will discuss further in Section~\ref{sec:discussion}.

\begin{figure}
    \centering
    \includegraphics[width=0.99\columnwidth]{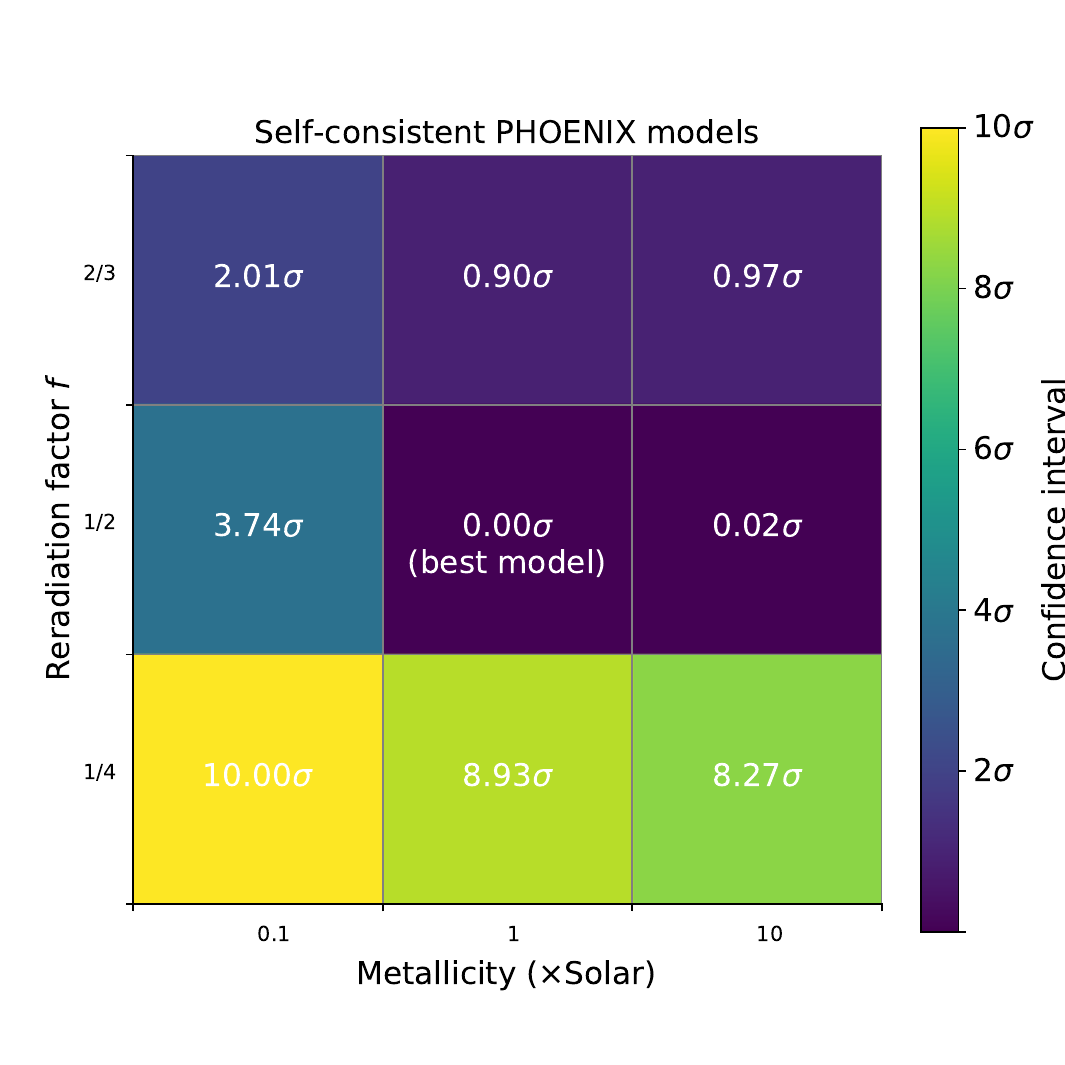}
    \caption{Confidence intervals for the PHOENIX self-consistent model grid for all observing nights combined using Wilks' Theorem. The best model, the one with the highest log-likelihood, is indicated. By definition of Wilks' Theorem, all confidence intervals here are with respect to this best model. The confidence intervals are converted to their corresponding $\sigma$-value (as described in the last paragraph of Section~\ref{s:cctologl}) which are annotated in the center of each tile. Models with $f \geq 1/2$ and metallicity $\geq 1 \times$ Solar are all within 1$\sigma$.}
    \label{fig:wilks_phoenix}
\end{figure}

While the modified PHOENIX models allow us to explore the structure and chemical species of the atmosphere of WASP-33 b, the self-consistent PHOENIX models allow us to assess its heat-redistribution efficient and  metallicity. The highest log-likelihood in the PHOENIX self-consistent model suite corresponds to a day-side-only heat-redistribution efficiency $f = 0.5$ and a metallicity of 1$\times$ Solar. Confidence intervals comparing the self-consistent PHOENIX models computed using Wilks' Theorem (with four free parameters) are shown in Fig.~\ref{fig:wilks_phoenix}, where we use a visualisation similar to that recently shown by \citet{Giacobbe2021}. Models with high re-radiation factors $f \geq 1/2$ and high metallicity $\geq 1 \times$ Solar are favoured and within 1$\sigma$ of our best model. We note that the higher S/N-ratio for the modified PHOENIX models compared to the self-consistent PHOENIX models is not necessarily surprising given the additional free parameters in the model and the larger number of models explored in the suite.

%

%
\subsection{Atmospheric dynamics and longitudinal variation}\label{sec:dynamics}
The post-eclipse phase coverage on Night 2 is longer than the pre-eclipse coverage on Nights 1 and 3. To enable a direct comparison, we split the post-eclipse Night 2 spectra into a subset of data that contains only the symmetric phases corresponding to the pre-eclipse coverage. A comparison of results from Night 2 and this subset of data with the different model suites are shown in Fig.~\ref{fig:corner_all} by the grey and green contours respectively, and in the right column of the S/N-maps in Fig.~\ref{fig:snr_main}. The S/N for the subset data is reduced, but only by $\Delta$S/N=0.7 at most, indicating that the data containing more night side hemisphere contributes less to the overall S/N of the detection. The green contours for the subset of phases are also tighter and better constrained in all cases in the $v_{sys}-K_{p}$ plots, again indicating a better match with the model when the night hemisphere contributions in the data set are reduced. As shown in Table~\ref{tab:snr_results_overview}, the scaling parameter from the CC-to-Log(L) mapping differs for the symmetric pre- and post-eclipse phases, consistent with a phase-dependence in the observed spectra of the planet due to asymmetry in the disk of the planet caused by e.g. an offset hot spot. To qualify this further, we use the GCM spectral templates described in Section~\ref{sec:gcm_modeling}.  

To first check for general consistency between the GCM and the results from the PHOENIX 1D models, we perform our two analysis methods using the day-side-only (i.e. no phase dependence) 3D GCM with all opacities included. This is detected at $S/N=7.1\sigma$, which is slightly lower the PHOENIX models, and found at the expected ($v_{\rm{sys}}$, $K_{\rm{p}}$)-values for all nights combined (see bottom row of Fig.~\ref{fig:snr_appendix} and the corresponding cross-correlation trail is shown in the middle row of Fig.~\ref{fig:gcm_trail_plots}). In terms of their likelihoods, we find a Bayes factor of +59.2 between the best PHOENIX 1D model and GCM suggesting the best PHOENIX 1D model is preferred over the GCM. However, we cannot compare the GCM and PHOENIX 1D models quantitatively in terms of their likelihood as they are drawn from different underlying parameter spaces. The preference of the best 1D PHOENIX model over the GCM is not necessarily surprising. This is because firstly, we explore only a single GCM day-side-only spectral template vs. many PHOENIX spectral templates, and secondly, the GCM lacks some of the important optical absorbers in its energy balance e.g. atomic iron. We can also assess in the day-side-only GCM model if any single species contributes the majority of the detection. The S/N results for single species models are shown in the right panel of Fig.~\ref{fig:snr_phoenix_modified_opacities} and the corresponding S/N-plots are shown in Fig.~\ref{fig:gcm_species}. From all opacity sources, only the CO GCM model results in a significant S/N of $6.6\sigma$. The GCM H$_{2}$O spectral template has $S/N=3.3\sigma$, which is insufficient for a robust detection in HRCCS as previous work demonstrated spurious signals at a S/N-ratio $\leq$4$\sigma$ often persist \citet{Cabot2020,Spring2022}, even though it appears at the expected planet velocity. OH is not detected. We also calculate the confidence intervals of each opacity source template to the case where all opacity sources are included. The results are shown in the right-panel of Figure~\ref{fig:opacities_wilks}. We find a model with only CO is significantly better ($\sigma=11.7$) compared to the day-side-only GCM with all opacity sources included, again in agreement with a robust detection of CO. Unsurprisingly, models with CO included are significantly better than models with just H$_2$O or OH. The CC-to-log(L) analysis shown in Fig.~\ref{fig:corner_opacity_models} further highlights the lack of robust detection of any species by CO with this model, in general agreement with the PHOENIX spectral templates. Although the low scale parameter for water in Fig.~\ref{fig:corner_opacity_models} could indicate water depletion from WASP-33 b's atmosphere, we find it more compelling that this results from telluric residuals in the final spectra, which is supported by the low S/N-ratio detection of the GCM H$_{2}$O spectral template and the large offset of this signal in ($v_{\rm{sys}}$, $K_{\rm{p}}$)-space away from the expected values for WASP-33 b.

Our main goal is to understand possible phase dependence in $\log(a)$. The bottom-left panel of Fig.~\ref{fig:corner_all} shows the results from the CC-to-log(L) framework for the GCM day-side-only model. The results are qualitatively similar to the the PHOENIX spectral templates, again with  offsets in $\log(a)$ for the different phase ranges, but with larger error margins. We hypothesize these larger error contours may be due to the broadening of the lines by planet rotation in the GCM spectral template and its overall lower S/N detection. The broad agreement with the trend for offsets in $\log(a)$ is confirmed by the day-side-only GCM and thus we proceed to allow the GCM to have phase dependence and determine if this resolves the $\log(a)$ scaling discrepancy.

The phase-dependent GCM provides a different cross-correlation template for each observed spectrum at each phase, where the spectrum corresponds to a P-T profile that a combination of all the profiles from different longitudes visible across the planet disk at that time. The phase-dependent GCM is similarly detected at $S/N=7.1\sigma$ at the expected ($v_{\rm{sys}}$, $K_{\rm{p}}$)-location for all night combined (the corresponding CCF-matrix and $K_{\rm{p}}-v_{\rm{sys}}$ maps are shown in the bottom rows of Fig.~\ref{fig:gcm_trail_plots} and Fig.~\ref{fig:snr_main}, respectively). However, the results from the CC-to-log(L) mapping shown in the bottom-right panel Fig.~\ref{fig:corner_all} are notably different for the scaling parameter. The phase-dependent GCM appears to resolve the majority of the $\log(a)$ discrepancy between the pre- and post-eclipse symmetric phases, and brings all the data sets into broad agreement within $2\sigma$. Although the phase-dependent GCM resolves most of the log(a) discrepancy seen for the 1D PHOENIX models, indicating that the inclusion of 3D dynamical effects such as hot spots and orbital phase brightness variation are a better match to the data, it does not fully describe the atmosphere of WASP-33 b. While the 1D model results in a higher S/N detection, it is possible that differences in full chemistry, or compensating with other parameters enable a better match, but only once scaled in the log-likelihood framework. The scaling parameter for the 1D models still indicates greater correction is needed for these models than the GCM. Thus there is scope for future work to improve on 3D models for ultra hot Jupiters to give a more comprehensive description of the atmosphere that would result in a higher S/N that the 1D models. This missing physics in the GCM may account for the overall lower S/N-ratio, compared to the S/N-ratio found for the best 1D PHOENIX models.

\section{Discussion}
\label{sec:discussion}
%
\subsection{Thermal structure and composition}
\label{sec:CO_detection}
The detection of CO emission lines with the modified PHOENIX P-T profiles provide unambiguous evidence for a thermal inversion in the atmosphere of WASP-33 b. A non-inverted atmosphere would have exhibited absorption lines instead. This is in agreement with previous observations of other atomic and molecular lines in emission in the atmosphere of WASP-33 b \citep[e.g.][]{Deming2012, Haynes2015, Zhang2018, Nugroho2020, Cont2021, Cont2021b, Herman2022} and expected from atmospheric modelling of UHJ atmospheres \citep[e.g.][]{Lothringer2018, Gandhi2019}. Although, we do not explore non-inverted or isothermal P-T profiles, our highest log-likelihood P-T profile (see Fig.~\ref{fig:pt_profle_confidence_intervals}) shows temperature contrasts with well constrained upper- and lower limits when including a thermal inversion layer.
From all molecular templates investigated in this work, only CO has resulted in a robust detection. This is further supported by the CC-to-log(L) results shown in Fig.~\ref{fig:corner_opacity_models} where only CO has similar constraints on the velocities compared to when all opacity sources are included: $v_{\rm{sys}} = 0.75^{+0.81}_{-0.79} \ \rm{km/s}$, $K_{\rm{p}} = 228.74^{+1.26}_{-1.29} \ \rm{km/s}$. Furthermore, only the exclusion of CO had a significant impact on the S/N-ratio of the planet signal. It does remain somewhat puzzling however that we still detect the best modified PHOENIX model without CO at a $\textrm{S/N} = 5.8$ in the first night of pre-eclipse data. We cannot attribute these to $\rm{H}_{2}O$ or OH alone in terms of the S/N-ratio, and we do not expect significant contributions from iron lines in the ARIES wavelength range either. However, if we compare the same models in terms of the CC-to-log(L) framework, we find exclusion of water or OH does result in a significantly worse model (see Fig.~\ref{fig:opacities_wilks}), perhaps suggesting a marginal detection of these species or the combined set of all other opacity sources, besides CO, during the first night.
However, we do notice that the confidence intervals calculated from Wilks' Theorem tend to suggest much higher $\sigma$-values than their S/N-ratio counterparts. Possibly, this is due to the assumption that the total number of spectral channels $N$ in equation~\ref{eq:logL_brogiline} all provide independent and uncorrelated measurements according to the definition of the likelihood by \citet{Brog19}. This assumption does not account for correlated noise between neighbouring pixels or for the fact that the FWHM of a resolution element may cover several pixels, particularly when observing at lower spectral resolutions. A full investigation into these effects on the CC-to-log(L) mapping and resulting $\sigma$-values is needed for this field of research.
We do not detect the model without CO during the third night, despite covering similar orbital phases. This may be explained by the better observing conditions during the first observing night relative to the third observing night (see S/N measurements in Table~\ref{tab:obs}). Although there have been many detections of CO in absorption \citep[][]{Sne10, Brog12, Rod12, deK13, Rod13, Lockwood2014, Brog14, Brog16}, this is the first  of CO emission lines using the HRCCS technique, using only the large Doppler-shift induced by the planet's orbital motion, that is without the aid of High Contrast Imaging.

Previous works have indicated both water \citep[][]{Haynes2015} and OH \citep[][]{Nugroho2021} in the atmosphere of WASP-33 b, but we do not make a robust detection of these molecules in the ARIES data. To demonstrate that the very weak GCM $\rm{H}_{2}O$ signal in our data is caused by non-planetary residuals e.g. tellurics, we ran the CC-to-log(L) analysis using only frames obtained when the planet is not visible i.e. during the full eclipse of WASP-33 b. Spurious signals due to telluric or stellar residuals should persist whereas signals originating from WASP-33 b should disappear or move away from the planet's expected ($v_{\rm{sys}}$, $K_{\rm{p}}$)-location. For $\rm{H}_{2}O$ we found that a signal is still retrieved at the same offset ($v_{\rm{sys}}$, $K_{\rm{p}}$)-position albeit with larger errors, while the CO signal was not obtained at the expected planet ($v_{\rm{sys}}$, $K_{\rm{p}}$)-position anymore. This supports the interpretation that any $\rm{H}_{2}O$ signal in the ARIES data is caused by non-planetary residuals.

The non-detection of OH in with ARIES/MMT is not necessarily surprising either when compared to the OH detection by \citet{Nugroho2021} using Subaru/IRD data. The bluer wavelength region (0.97-1.75 $\micron$) covered by Subaru/IRD with respect to ARIES/MMT contains stronger OH emission lines and Subaru/IRD's has higher spectral resolution ($R = 70,000$). Thus, the lack of an OH detection with ARIES can likely be explained by the poorer data quality of the spectral orders covering the OH line dense regions compared to CO line dense regions. The synthetic GCM spectrum in the left-top panel of Fig.~\ref{fig:gcm_models_overview} shows OH line dense regions in the wavelength range of the bluer spectral orders and order 26 at the far red-end. As explained in Section~\ref{s:sn-method}, lower weights are assigned to the bluer ARIES orders which have a lower throughput and contain residual fringing. The reddest order 26 falls at the edge of the detector and is consequently of poorer data quality. Lastly, telluric absorption is strong in some OH line dense spectral orders.

Due to the high dissociation temperature for CO, the abundances of CO are approximately constant in- and outside of the hotter and cooler regions of the planet, at the pressures probed by HRCCS \citep{Lodders2002}.
This is in line with modeling results of WASP-33 b by \citet{Tsai2021} who also found CO to be the only molecule that can be stable against dissociation. This is in contrast to other molecular species such as $\rm{OH}$, $\rm{H}_{2}O$ or TiO which dissociate on the day-side (at least up to temperatures of $\sim$4000 K at a pressure of $10^{-4}$ bar). For example, for $\rm{H}_{2}O$ this will result in an abundance difference of several orders of magnitude between the day- and night-side. Since we assume Solar metallicity, and the Solar C/O value is $\sim$0.54 \citep{Asplund2009, Caffau2011}, the CO abundance will not really depend on the varying H$_2$O abundance, but rather the limiting factor to form more CO will be the amount of carbon-atoms available. This is regardless of the extra freed-up oxygen atoms that become available as water dissociates at higher day-side temperatures. For species other than CO, their abundance has a strong temperature-dependence and introduces a degeneracy between the effects of the atmospheric abundance (or the dissociation rate) and thermal structure, complicating the interpretation of the line contrast. This stability of CO against dissociation means the CO line contrast is a reliable tracer of the thermal structure of the atmosphere.
We therefore advocate for CO as a better probe of the thermal structure and inversions of UHJs compared to $\rm{H}_{2}O$ in HRCCS.

\subsection{Robustness of CO as a temperature tracer in the presence of stellar activity}
The CCFs inside the planet radial velocity trail in Fig.~\ref{fig:phoenix_cc_trail} are distinct and easily seen even by eye. We do not see contaminating signatures of the $\delta$ Scuti pulsations. On the contrary, the contamination by stellar pulsations can be seen clearly in the cross-correlation analyses neutral iron emission lines \citep{Nugroho2020a, Herman2022}. This is not exclusive to WASP-33 b; any pulsating star with opacity sources present in both the star and planet will contaminate HRCCS. This contamination limits the available effective orbital phase coverage, as frames obtained at phases close to secondary-eclipse have to be either heavily processed to remove the stellar pulsations \citep[][]{Johnson2015, Temple2017, vanSluijs2019} or disregarded \citep{Nugroho2020a, Herman2022,Spring2022}. In contrast to atomic species often present in both the planet and star, CO is dissociated in the hot stellar atmosphere of early-type A/F-stars. These type of stars are also more likely to be pulsating stars due to their location within the Hertzprung-Russel diagram's instability strip \citep[e.g.][]{Gautschy1996}. This highlights a further advantage of using CO in the NIR to probe the atmospheric temperature structure and inversions of UHJ around hot pulsating stars.
\subsection{Atmospheric dynamics and longitudinal variation}
\label{sec:eastward_hotspot}
\begin{figure*}
    \centering
    \includegraphics[width=0.75\textwidth]{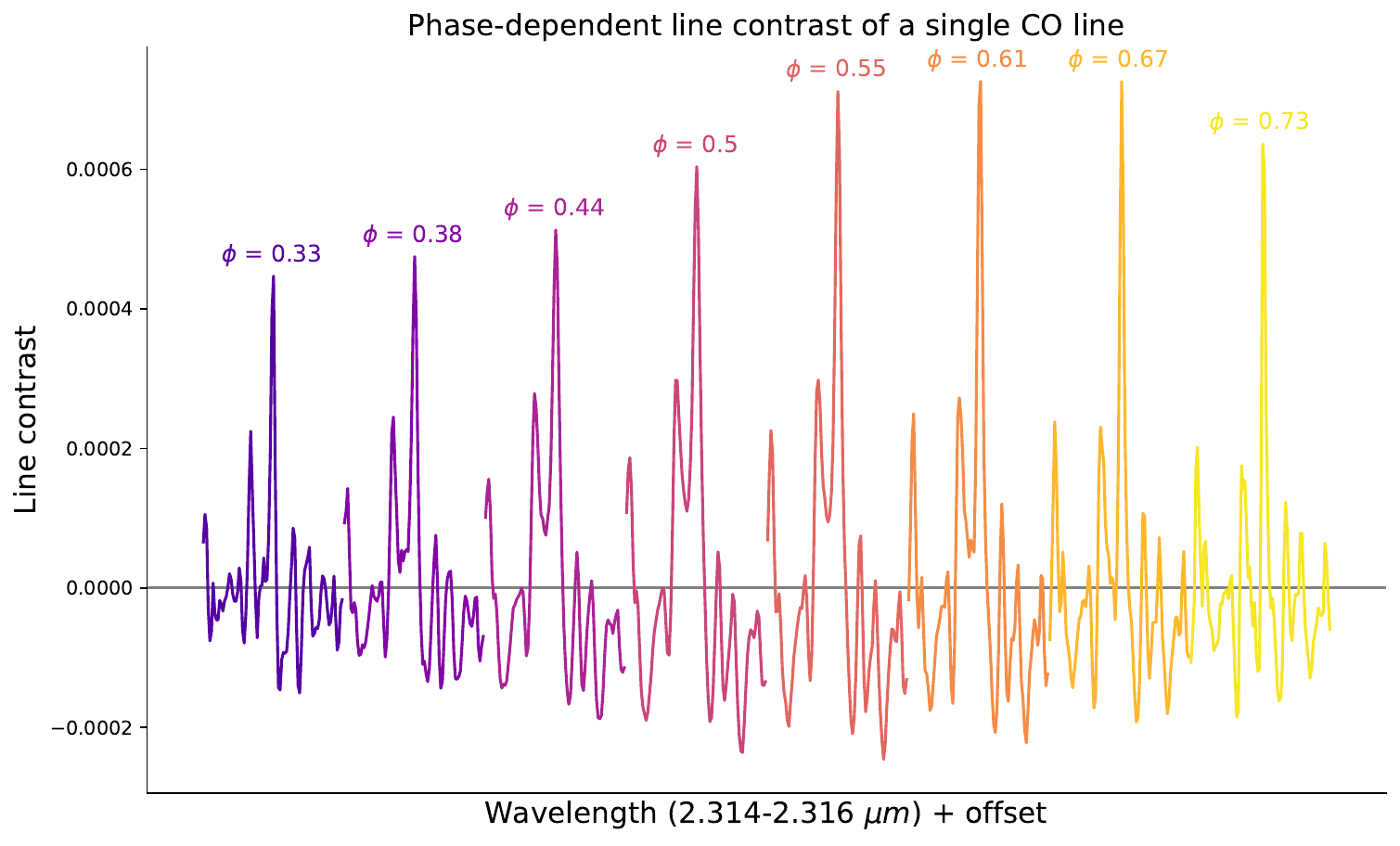}
    \caption{Line contrast (flux - mean(flux)) of a single CO line (2.314-2.316 $\micron$) in the phase-dependent GCM as function of orbital phase ($\phi$) coverage. The phase-dependent variations show shallower lines pre-eclipse compared to post-eclipse for symmetric phases around the secondary eclipse. This is due to a shallower average temperature gradient in the pre-eclipse phases compared to post-eclipse.}
    \label{fig:line_contrast}
\end{figure*}

Hydrodynamical simulations predict the thermal structure of hot Jupiters to be asymmetric around the sub-stellar point, with the hottest point being shifted eastward. This shift arises from the competition between heat transport by winds of planetary-scale waves and the radiative cooling of the parcel of gas \citep[e.g.][]{Showman2002, Perez-Becker2013}.
As the exoplanet orbits and its host star, its synchronous rotation causes more of its eastward side to be visible pre-eclipse, and more of its westward side post-eclipse. Consequently, if an eastward hotspot exists, the pre-eclipse phases are more dominated by the hot spot than the post-eclipse phases.

Previous evidence for a hot spot on WASP-33 b comes from NIR and optical photometric phase curve observations, where eastward hot spot results in negative phase curve offset. \citet{Zhang2018} report the first evidence of an eastward hot spot on WASP-33 b, measuring a phase curve offset of $-12.8 \pm 5.8 \degree$ in the Spitzer $3.6 \micron$-band. Conversely, \citet{vonEssen2020} found a $+28.7 \pm 7.1 \degree$ westward phase curve offset in the optical TESS light curve, and a range of theories including magnetohydrodynamic effects, non-synchronous rotation, and clouds have been invoked to explain such westward offset hot spots \citep[e.g.][]{Dang2018,Hindle2021}. Recent work by \citet{Herman2022} further report a $+22 \pm 8 \degree$ westward offset based on the phase offset of neutral iron thermal emission lines at high spectral resolution. Our CO emission line detection also supports a phase-dependence in the planet spectrum. However, the GCM that we use to model it includes an eastward (not westward) hot spot, which is oriented towards us during the pre-eclipse phases of WASP-33 b.

It is important to note that the phase-dependence detected in this work and in \citet{Herman2022} is found via a model scaling parameter ($a$ in this work, and a similar $A_{p}(\phi)$ in equation 4 of the other study). Both studies detect that a larger scaling is required to best model the observations just after eclipse. \citet{Herman2022} interpret the larger $A_{p}(\phi)$ post-eclipse as an overall brightness variation, indicating more flux west from the sub-stellar point and hence a westward hot spot. This is the typical interpretation for photometric phase curve measurements, which are sensitive to absolute flux as a function of phase. Instead, we highlight that because the HRCCS we use in this work divides out the continuum, that the scaling parameter is a measure only of line contrast i.e. that the lines in the planet spectrum are deeper just after eclipse, as shown in Figure~\ref{fig:line_contrast}. 

The phase-dependent line contrast can then be understood in the context of our GCM spectral template. First, the eastward hot spot results in a hotter continuum and thus greater overall brightness for pre-eclipse phases as shown in the bottom-left panel of Fig.~\ref{fig:gcm_models_overview}. But, since our post-processing of the high resolution spectra removes the continuum information we argue we are insensitive to these variations. Fig.~\ref{fig:gcm_models_overview} then shows that the GCM predicts thermal profiles that are shallower in the eastward-shifted hot spot and steeper on the western part of the day-side. Overall, this leads to a disk-integrated temperature gradient that is shallower during pre-eclipse when it contains the eastward hot spot (where our line-of-sight is centered on longitudes $\sim17-61 \degree$ east of the sub-stellar point), and steeper during post-eclipse when it contains more of the western day-side hemisphere (where our line-of-sight is centered on longitudes $\sim277-341 \degree$ east of the sub-stellar point). Consequently, the average line contrast associated with these temperature gradients will be shallower during pre-eclipse and stronger during post-eclipse. For a fixed line contrast amongst all phases, such as the PHOENIX models or the day-side-only GCM, this results in a smaller scale parameter pre-eclipse compared to post-eclipse, as observed in our dataset. But for the 3D phase-dependent model, this results in a more consistent scale parameter between the pre- and post-eclipse data, again as observed in our dataset. We therefore conclude that despite our insensitivity to the absolute brightness variations of the hot spot, the phase-dependent variation of the scale parameter is still sensitive to the 3D nature of the atmosphere and consistent with an eastward offset hot spot scenario.
As a check, we also investigate if this result is robust against the alternative scenario of a westward offset hot spot. Such a scenario is not predicted by our GCM, but might arise from magnetohydrodynamical effects on the atmospheric dynamics \citep{Hindle2021}. Including magnetohydrodynamical physics into the GCM is beyond the scope of this work. Instead we simply mirror our phase-dependent GCM model such that eastward phases becomes westward and vice versa, effectively turning our eastward offset model into a westward offset model. The result is shown in Figure~\ref{fig:corner_westward}. The discrepancy between pre- and post eclipse phases has significantly increased ($\Delta \log(a) = 0.41^{+0.03}_{-0.04}$) due to the even shallower line contrast post-eclipse compared to pre-eclipse. We conclude a model with a westward hot spot offset is inconsistent with our data, but that the assymmetry therefore is consistent with an eastward offset.
The lower S/N-ratio of the phase-dependent GCM compared to the PHOENIX models may seem unexpected at first, also in the light of results from \citet{Beltz2021} who find a significant increase of the S/N-detections for the exoplanet HD 209458b when using a 3D GCM compared to a 1D model. In part this may be explained by the lower spectral resolution of this data compared to \citet{Beltz2021} as the 3D effects of the line shape and Doppler shift of the spectral lines will be more pronounced at higher spectral resolutions where they can be resolved. We also compare a single GCM (N=1) to a suite of PHOENIX models (N=576 modified models and N=9 self-consistent models), which is an unfair comparison. It may be explained current limitations of the GCM, mainly the exclusion of Fe and Fe$^+$. \citet{Lothringer2018} find Fe and Fe$^+$ are important opacity sources in the atmospheres of UHJ that contribute to their thermal inversions. Nonetheless, we emphasize the this phase-dependent change of P-T profile with longitude is robust to the choice inclusion or exclusion of iron opacities in our GCM. When an eastward hot spot shift is present, air west of the hot spot shift is experiencing a transition from cold to hot. This transition happens first at low pressures and propagates downwards. This naturally makes a very strong thermal gradient at this transition between the cold and hot profile and happens at the pressure these observations probe. Once air flows from the hot spot to the cold side the cooling happens more homogeneously over the P-T profile, leading to a less steep gradient. This behaviour does not depend on the details of the hot and cold profiles and should be the same regardless of whether iron opacities were included or not. Instead the phase-dependent variation of the thermal gradient and thus line contrast should be seen as a natural outcome of having a shifted hot spot.


%
We do not find significant evidence for a relative blue-shift pre-eclipse compared to post-eclipse that would further indicate an eastward hot spot. Such net Doppler shifts are predicted of $\sim$1-3 km/s as the hot spot rotates in and out of the observers view over while orbiting its host star \citep{Zhang2017, Beltz2021}. However, our velocity sampling is $\sim$4-6 km/s for the individual observing nights. Thus we cannot resolve such Doppler shifts at our spectral resolution.
We note that when including all orbital phases out to quadrature in the post-eclipse night, the scaling parameter prefers shallower lines (see grey contours in Fig.~\ref{fig:corner_all}), despite the strong line contrasts shown in Fig.~\ref{fig:line_contrast}. While these later phases add only a small contribution to the overall S/N, the change in line contrast may be in part explained by the post-processing of the data. The rate of change of the radial velocity for close-to-quadrature phases is small, and therefore our PCA-cleaning will be more effective in removal of the exoplanetary signal, and this has also been reported by \citet{Brogi2022} and \citet{Pino2022}. We thus emphasize it is key to only compare symmetric phases, ensuring both the pre- and post-eclipse frames included were homogeneously processed by our post-processing steps.

We further find a systematic bias towards small scaling parameter $\log{a} < 0$ in all models. Rather than missing physics, we consider that the data cleaning steps of the HRCCS can affect the line shape and strength of planetary signal. PCA likely degrades more of the planetary signal at lower resolution, where the lines spread across multiple pixels. Applying PCA to our template in every iteration of our CC-to-log(L) {\sc PyMultinest} implementation is computationally expensive. As an alternative, we investigated these effects by injecting the best modified PHOENIX model at $a=1$ into the frames observed during the eclipse of WASP-33 b. Since the change in radial velocity would be relatively large during the eclipse, potentially making it more robust against PCA, we shifted the injected model in phase to cover similar radial velocities as observed during the first observing night. We found a scale parameter of $\log a = -0.23^{+0.03}_{-0.03}$ for the recovered signal. This result supports our hypothesis that our data cleaning steps cause at least part of the bias towards lower scale factors. Given that we apply the data processing in a homogeneous way to all data sets, and while we do not give large weight to the absolute values of $\log(a)$, we still give weight to the interpretation of the relative differences.

%
Future work adopting novel model-filtering techniques such as the on introduced by \citet{Gibson2022} may resolve the issue of comparing asymmetric phase coverage pre- and post-eclipse and alleviate the systematic bias towards lower scaling parameters.

\subsection{Low resolution limit for HRCCS technique}
\label{sec:lowestres_detection}
For the HRCCS technique, to the first order, the S/N-ratio of the planet $S/N_{\rm{p}}$ is given by
\begin{equation}
    {S/N_{\rm{p}} = \frac{S_{\rm{p}} \sqrt{N_{\rm{lines}}}}{\sqrt{S_{\rm{\star}} + \sigma_{\rm{bg}}^2 + \sigma_{\rm{read}}^2 + \sigma_{\rm{dark}}^2}}},
    \label{eq:snrplanet}
\end{equation}
where $S_{\rm{p}}$ is the planet's flux, $S_{\rm{\star}}$ the stellar flux, $N_{\rm{lines}}$ the number of resolved spectral lines, $\sigma_{\rm{bg}}$ the noise from the sky and telescope background, $\sigma_{\rm{read}}$ the detector read-out noise and $\sigma_{\rm{dark}}$ the detector dark current \citep{Snellen2015}.
In the photon-limited regime, Equation~\ref{eq:snrplanet} reduces to
\begin{equation}
    {S/N_{\rm{p}} = (\frac{S_{\rm{p}}}{S_{\rm{\star}}}) \ S/N_{\rm{\star}} \ \sqrt{N_{\rm{lines}}}},
\end{equation}
where $\textrm{S/N}_{\rm{\star}}$ is the stellar S/N-ratio \citep{Birkby2018}. Our observations are photon-limited thanks to the bright host star. However, residual noise, likely from residual telluric lines, persists in the final CCF time series (see Fig.~\ref{fig:phoenix_cc_trail}), but the impact of this is substantially mitigated by the large Doppler shift of WASP-33 b.
For the close-in UHJs, we cannot be aided by High Contrast Imaging to spatially resolve the exoplanet and increase our planet-to-star contrast ratio. We thus rely solely on disentangling the exoplanet's lines from it's the Doppler shift induced by its orbital motion. We can thus increase our exoplanet's S/N in two ways: Firstly, we can increase our photon collecting power such that the $\textrm{S/N}_{\rm{\star}}$ increases or secondly increase the total number of spectral lines observed by either expanding our instantaneous wavelength coverage or our spectral resolving power.
Since S/N is proportional to $\sqrt{N_{\rm{lines}}}$ it is crucial to have a sufficiently high spectral resolution to resolve more lines. Previously, the lowest spectral resolution at which HRCCS unaided by high contrast imaging was successfully demonstrated was $R=25,000$, as shown by the detection of water vapour in the atmosphere of $\tau$ B\"oo with the pre-upgraded NIRSPEC1.0/Keck combination as reported by \citep{Lockwood2014}. Other evidence for a detection of the thermal spectrum of HD 88133 b, also with NIRSPEC1.0/Keck was reported by \citet{Piskorz2016} using a multi-epoch approach, but later on disputed \citep{Buzard2021}. Our detection with ARIES/MMT shows the Doppler-only HRCCS can still be used even at spectral resolutions as low as $R = 14,000-16,000$.
Our detection of WASP-33 b at this lower spectral resolution provides proof-of-concept HRCCS is still possible at a lower resolving power when there is a sufficiently large orbital Doppler shift, despite less and diluted spectral lines being observed. Therefore, one can potentially trade off spectral resolving power for photon collecting power. This is particularly interesting for short period exoplanets where we do not need a high spectral resolving power to resolve their large orbital Doppler shifts. Examples include other UHJs (for example KELT-9b) or terrestrial lava worlds (for example CoRoT-7b) as both have orbital velocities of similar order of magnitude as WASP-33 b.

\section{Conclusions}
\label{sec:conclusions}
In this work we presented the first high resolution spectroscopy observations from the NIR spectrograph ARIES mounted behind the MMT targeting WASP-33 b. We develop a public pre-processing pipeline for ARIES to extract a spectral time series from the raw detector images. We use the HRCCS technique with PCA to characterise WASP-33b's atmosphere using both the S/N-method and CC-to-log(L) mapping. The primary conclusions of this work are the following:

\begin{enumerate}
    \item We present the first robust detection (7.9$\sigma$) of CO emission lines using HRCCS in an exoplanet atmosphere. The detection of CO emission lines in the atmosphere of WASP-33 b is unambiguous evidence of an inverted P-T profile.
    \item The best self-consistent PHOENIX  atmospheric models prefer day-side-only heat redistribution.
    \item The modified P-T profile PHOENIX models show we are sensitive to the temperature gradient of the inversion layer, but less sensitive to the absolute temperatures of the lower atmosphere.
    \item The CC-to-log(L) mapping reveals a phase-dependent scaling factor, which shows that the planet spectral line contrast is greater just after secondary eclipse than before. We demonstrate that the P-T profiles from a phase-dependent GCM that includes an eastward offset hot spot can explain the majority of the phase-dependence in $\log(a)$. We emphasise that inclusion of a scaling parameter in the CC-to-log(L) mapping can reveal the longitudinal phase-dependent thermal structure of the atmosphere, even in spectra that lack the resolving power required to detect the additional induced Doppler shift caused by the offset hot spot.
    \item We find no evidence of stellar pulsations affecting our detection of CO, in contrast to previous detections of atomic emission lines in the optical. We advocate that CO is therefore an advantageous and good probe of the thermal atmospheric structure of UHJs at pressures seen with HRCCS.
    \item At a measured resolving power of $R\sim15,000$, these observations show that HRCCS can be used even at spectral resolutions of $R\sim15,000$ when using only the large Doppler-shift induced by the exoplanet's orbital motion to disentangle its spectrum from its host star and tellurics. That is without the use of any spatial information from High Contrast Imaging. Thus spectral resolving power can be traded for photon collecting power in cases where the induced orbital Doppler shift is sufficient move the planet spectrum over multiple pixels such as expected for close-in terrestrial lava planets and ultra hot Jupiters.
\end{enumerate}

ARIES is currently being upgraded into the MMT: AO exoPlanet characterization System (MAPS)\footnote{More information can be found here: \url{https://www.as.arizona.edu/~ktmorz/maps.html}}. It will have an updated AO system, broader instantaneous wavelength coverage and modes at a higher spectral resolution. A modified version of the ARIES data reduction pipeline presented in this work may prove helpful in analysing future output from MAPS as well.

Finally, future low-resolution observations of WASP-33 b with the James Webb Space Telescope can be combined to with ground-based high-resolution observations to mitigate the effects of stellar pulsations and give tighter constraints on the P-T profile and chemical abundances of $\rm{OH}$, $\rm{H}_{2}O$, $\rm{CO}$, and $\rm{CO}_{2}$ \citep[][]{Beichman2018}. UHJ have the additional benefit that refractory elements can be measured as well \citep[][]{Lothringer2021}. Combined this will enable calculation of abundance ratios which can in turn be used to constrain different planet formation pathways \citep[e.g.][]{Oberg2011,Madhusudhan2012,Cridland2016,Cridland2017,Cridland2019,Khorshid2021}. Finally, combining phase-dependent low resolution data, for example the recently observed diurnal variations in the atmosphere of WASP-121b \citep[][]{Mikal-Evans2022}, and high resolution data using the CC-to-log(L) \citep[][e.g.]{Brog17,Brog19,Gibson2021} mapping will greatly improve our understanding of the longitudinal chemical and thermal phase-dependence.


\section*{Acknowledgements}
We thank the anonymous reviewer for the careful reading of our manuscript and the many insightful comments and suggestions. JLB extends special thanks to all the MMT Observatory staff, particularly the telescope and AO operators, for their assistance and valuable advice in carrying out the MMT Exoplanet Atmosphere Survey. We thank K. Powell for all his assistance with the AO instrumentation operation during the process of collecting these data. This research is part of the exoZoo project that has received funding from the European Research Council (ERC) under the European Union's Horizon 2020 research and innovation program under grant agreement No 805445. Observations reported here were obtained at the MMT Observatory, a joint facility of the Smithsonian Institution and the University of Arizona. This work was performed in part under contract with the Jet Propulsion Laboratory (JPL) funded by NASA through the Sagan Fellowship Program executed by the NASA Exoplanet Science Institute. M.B. acknowledges support from the STFC research grant ST/T000406/1.

\section*{Data Availability}
All the original data and pre-processing pipeline is publicly available in the author's Github repository: \url{https://github.com/lennartvansluijs/ARIES}.



\bibliographystyle{mnras}
\bibliography{referencesjlb} 




\appendix

\section{Additional tables and figures}
For transparency, completeness and reproducibility of this work, we include some additional figures and tables here.

\begin{table*}
    \centering
    \begin{tabular}{lccr}
    \hline
        Quantity & Value & Unit & Reference \\
    \hline\hline
        Stellar mass & 1.495 & $\rm{M}_{\odot}$ & \citet{Cameron2010} \\
        Stellar effective temperature & 7400 & K & \citet{Cameron2010} \\
        Stellar radius & 1.444 & $\rm{R}_{\sun}$ & \citet{Cameron2010} \\
    \hline
        Primary transit time & 2454590.17936 & BJD & \citet{Smith2011} \\
        Eccentricity & 0.0 & - & \citet{Smith2011} \\
        Semi-major axis & 0.02558 & au & \citet{Smith2011} \\
        Orbital period & 1.21986967 & days & \citet{Smith2011} \\
        Orbital inclination & 87.7 & degree & \citet{Lehmann2015} \\
        Planetary radius & 1.603 & $\rm{R}_{\rm{Jup}}$ & \citet{Lehmann2015} \\
        Impact parameter & 0.21 & - & \citet{Chakrabarty2019} \\
    \hline
        Expected system velocity & -0.3 & km/s & \citet{Nugroho2020} \\
        Expected orbital velocity & 230.9 & km/s & \citet{Nugroho2020} \\
    \hline
    \end{tabular}
    \caption{An overview of relevant WASP-33 system parameters used in this work.}
    \label{tab:wasp33_system}
\end{table*}
\begin{figure*}
    \centering
    \includegraphics[width=\textwidth]{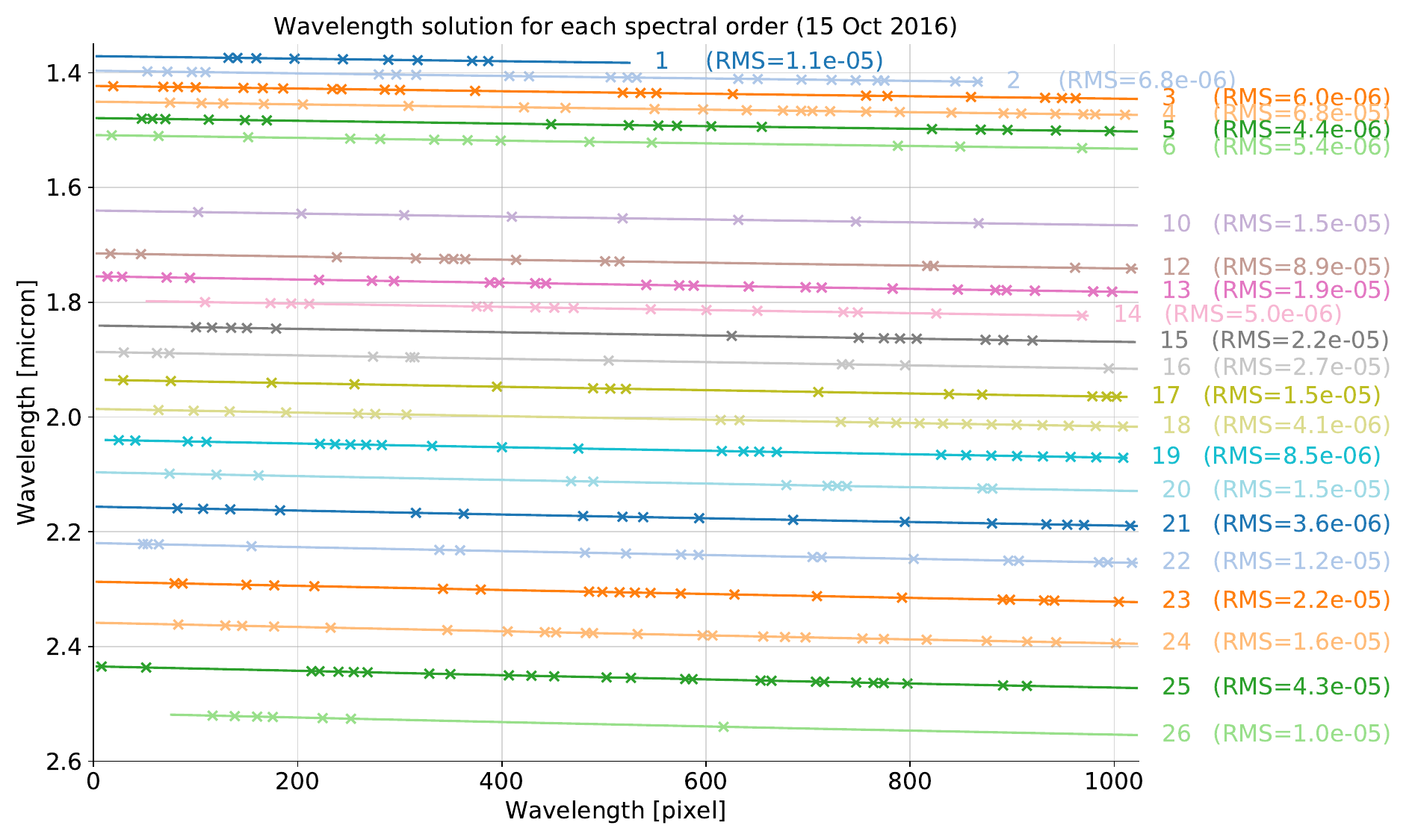}
    \caption{Wavelength solution for each spectral order for the first observing night. For each order the points indicate visually identified telluric absorption lines and the lines indicate the best-fit polynomial. The root mean square (RMS) values of the difference between the points and the wavelength solution are noted to indicate quality of the fit for each spectral order. Some masked orders for which no good wavelength solution was found have been omitted.}
    \label{fig:wavsolution_15oct2016}
\end{figure*}
\begin{figure*}
    \centering
    \includegraphics[width=\textwidth]{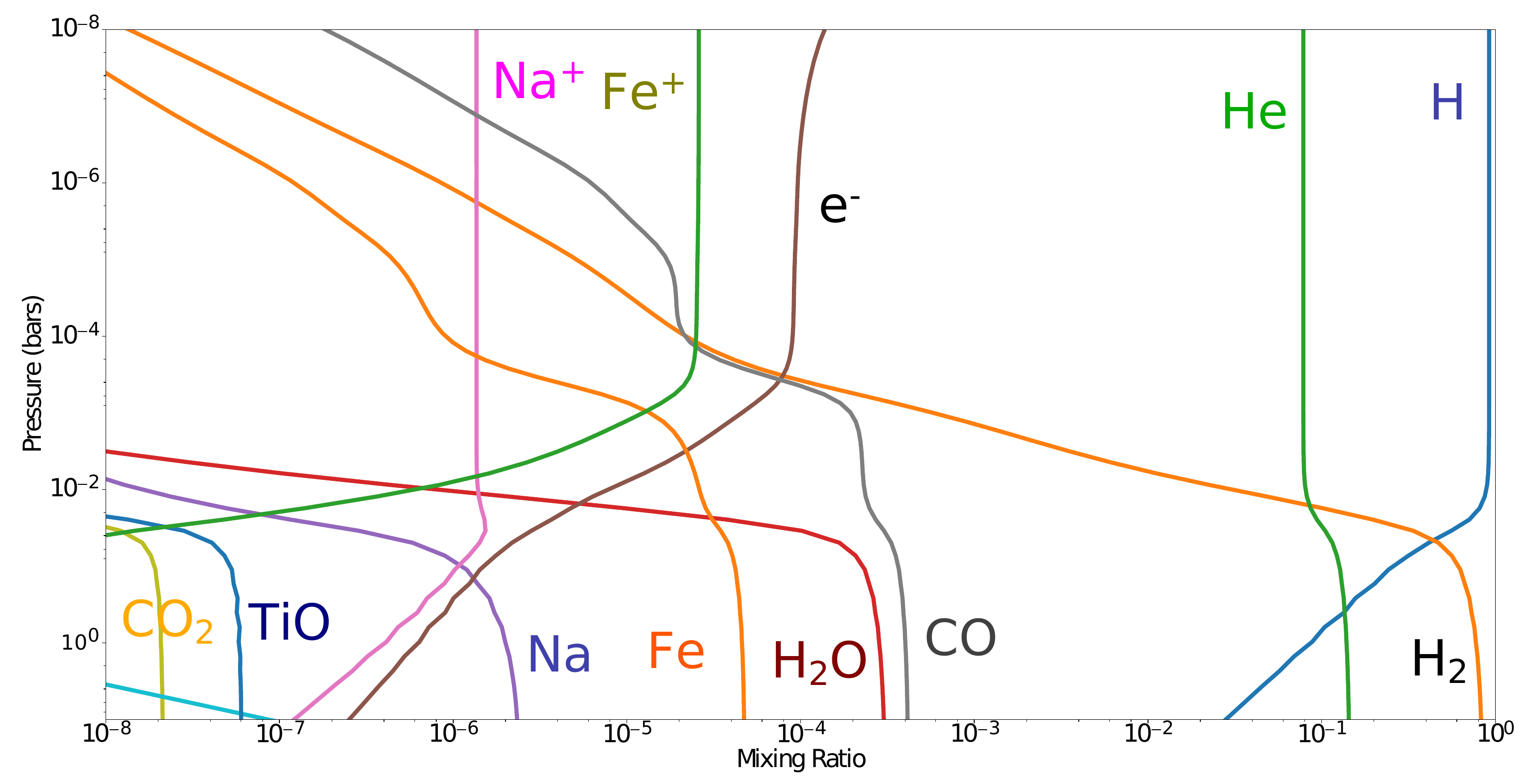}
    \caption{Mixing ratios in chemical equilibrium as functions of pressure for different species in the self-consistent PHOENIX model atmospheres of WASP-33 b. This is for the day-side-only heat redistribution model $f=0.5$ at $\times 1$ Solar metallicity.}
    \label{fig:phoenix_mixing_ratios}
\end{figure*}
\begin{figure*}
    \centering
    \includegraphics[width=0.8\textwidth]{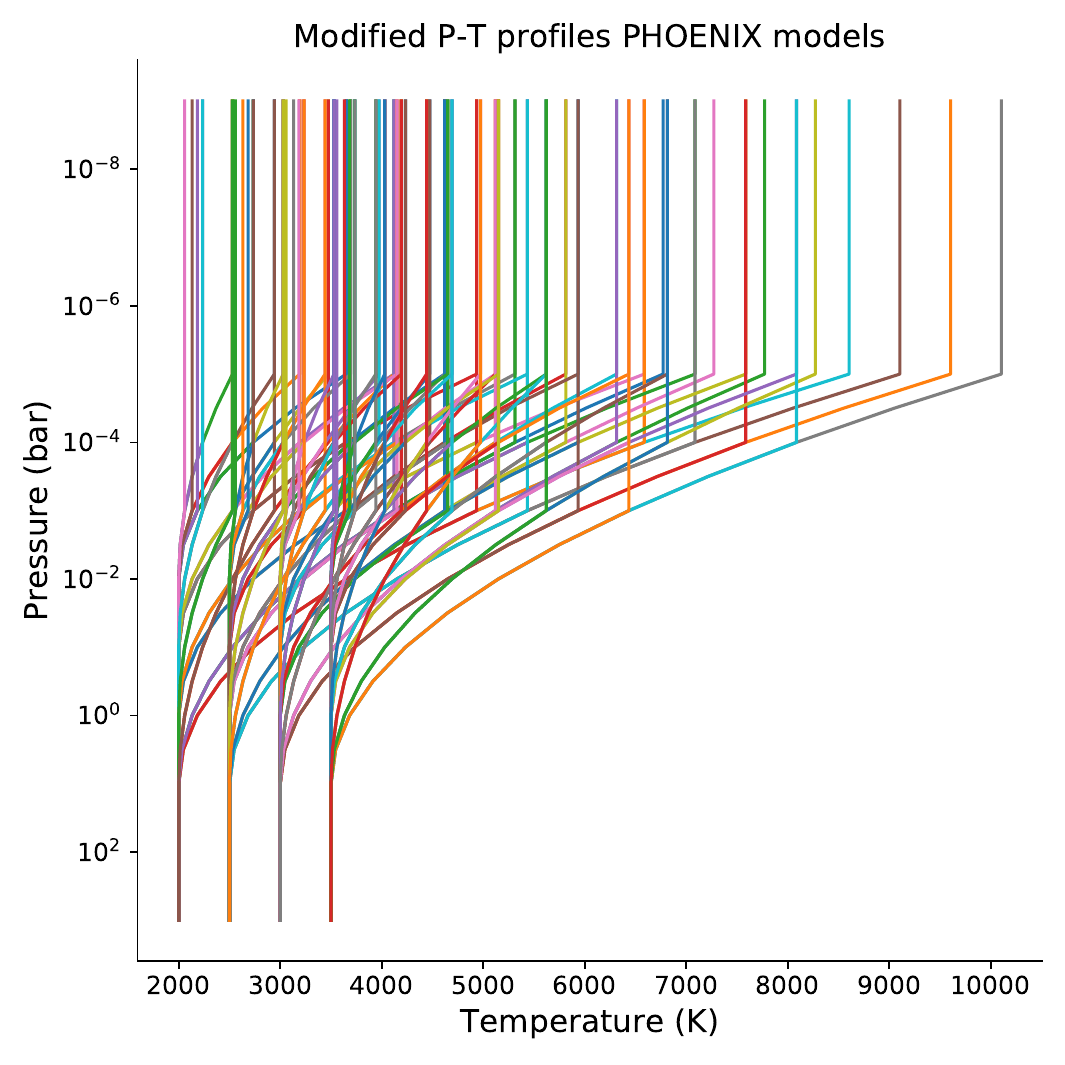}
    \caption{Overview of all modified PHOENIX P-T profiles explored in this work.}
    \label{fig:phoenix_modified_structures}
\end{figure*}
\begin{figure*}
    \centering
    \includegraphics[width=0.95\textwidth]{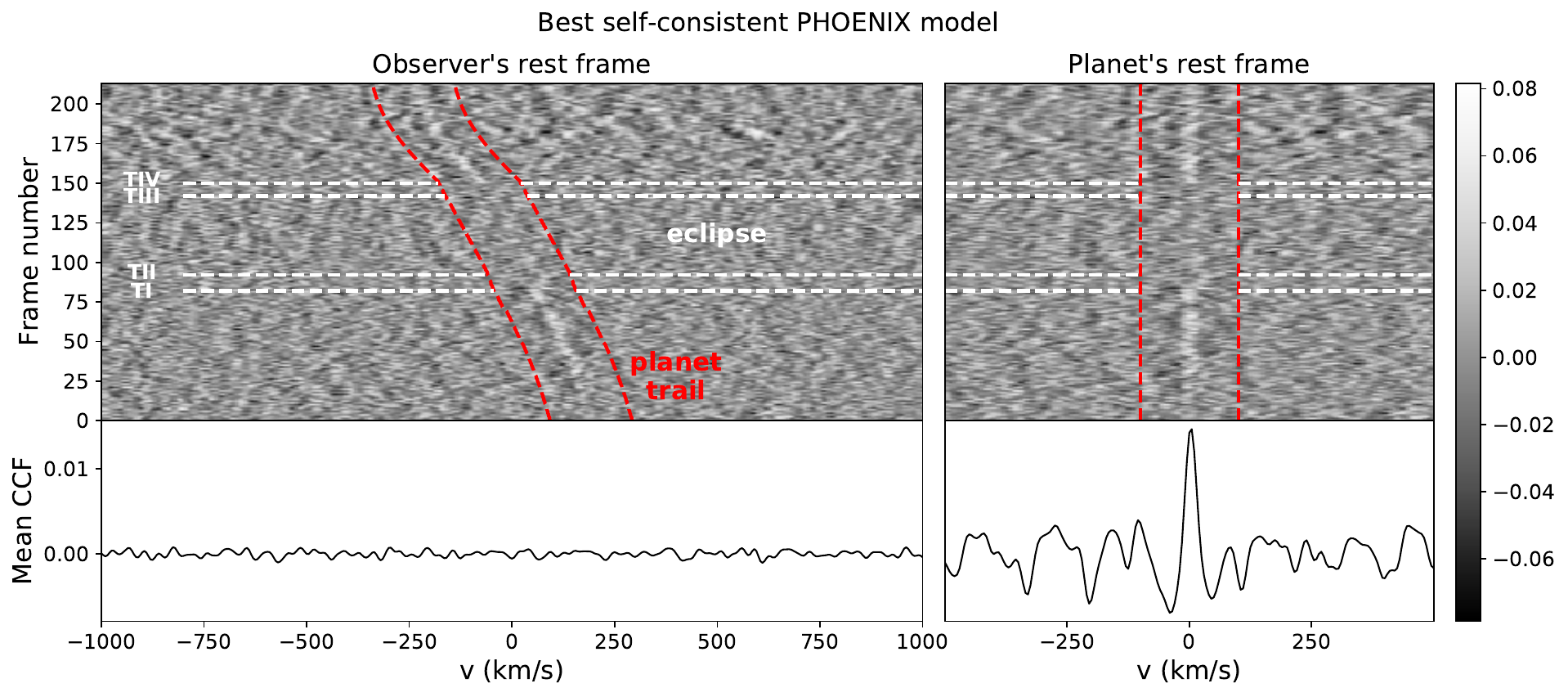}
    \includegraphics[width=0.95\textwidth]{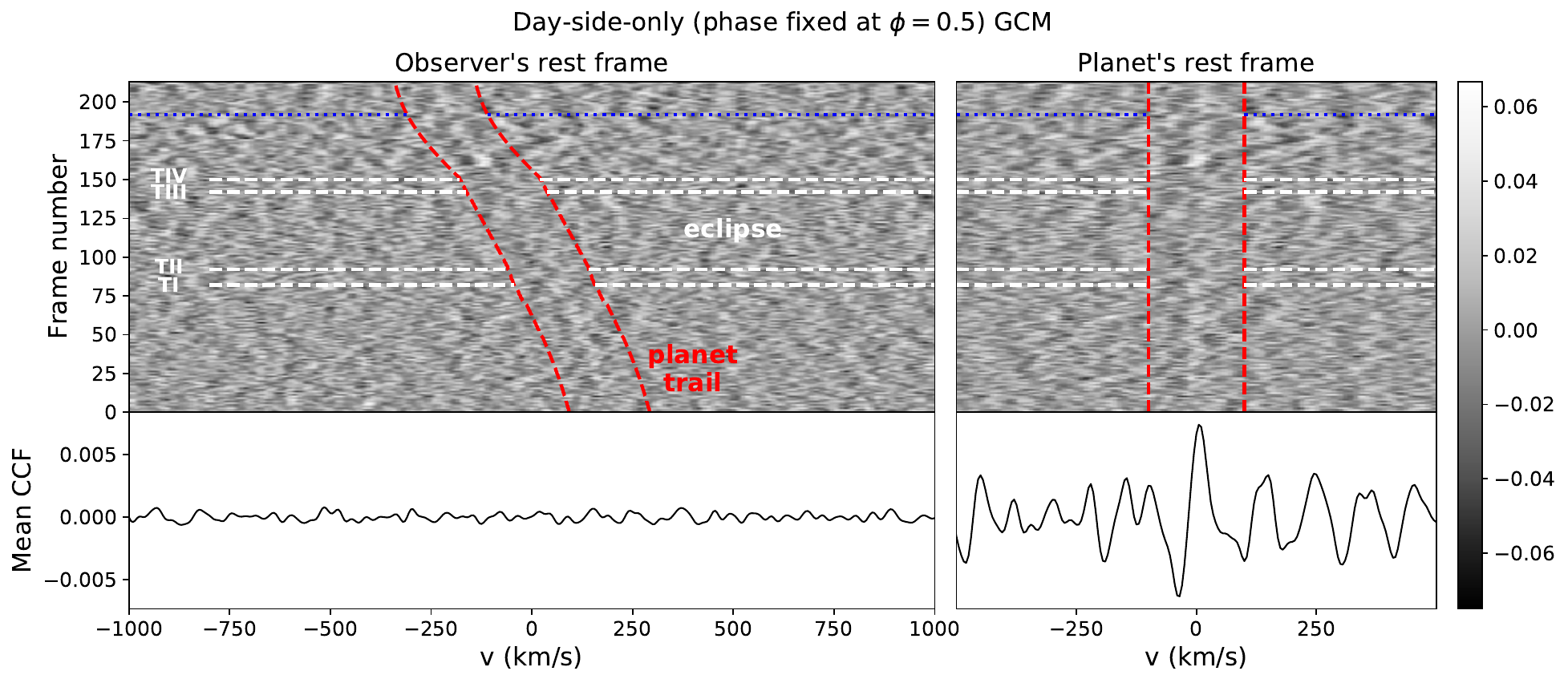}
    \includegraphics[width=0.95\textwidth]{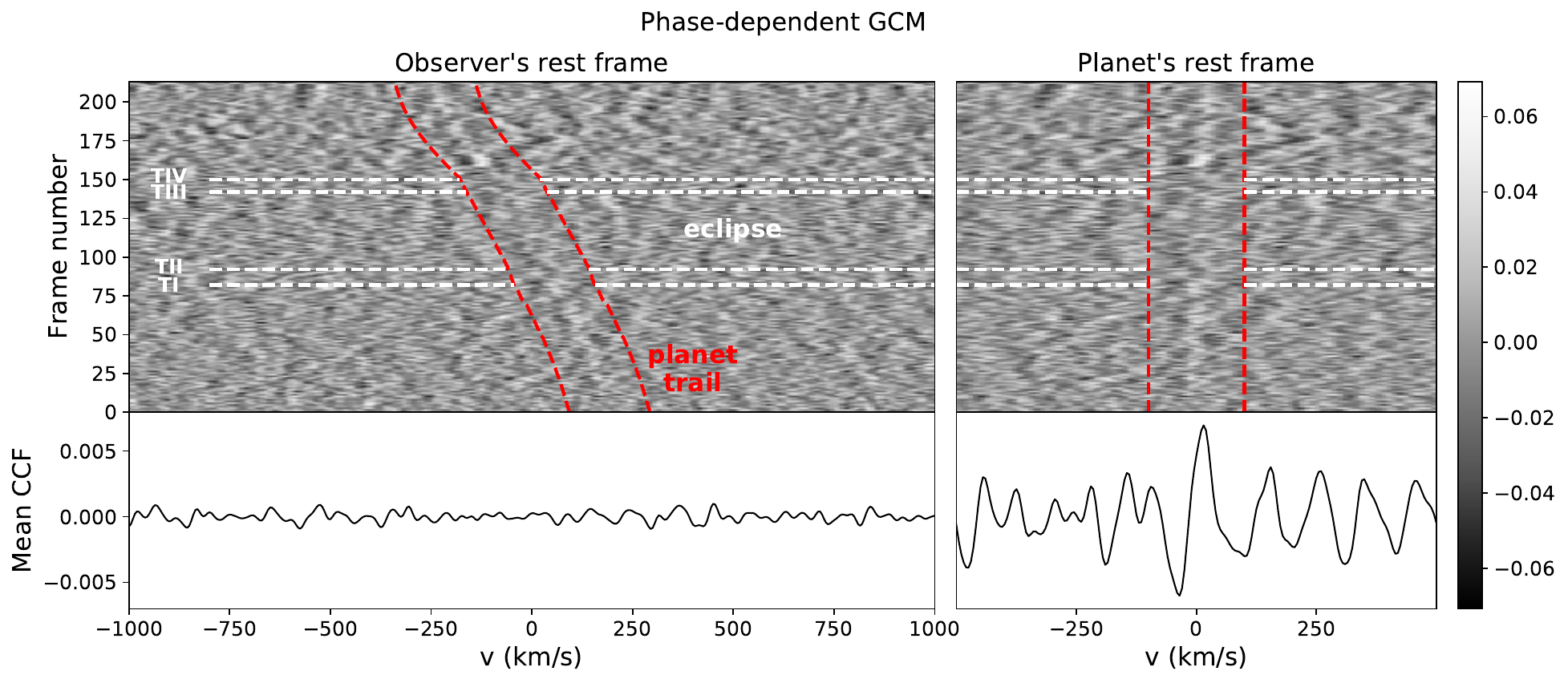}
\caption{Similar to Fig.~\ref{fig:phoenix_cc_trail}, but for the highest S/N-ratio self-consistent PHOENIX model, day-side-only GCM (middle panel) and phase-dependent GCM (bottom panel).}
    \label{fig:gcm_trail_plots}
\end{figure*}
\begin{figure*}
    \centering
    \includegraphics[width=0.5\textwidth]{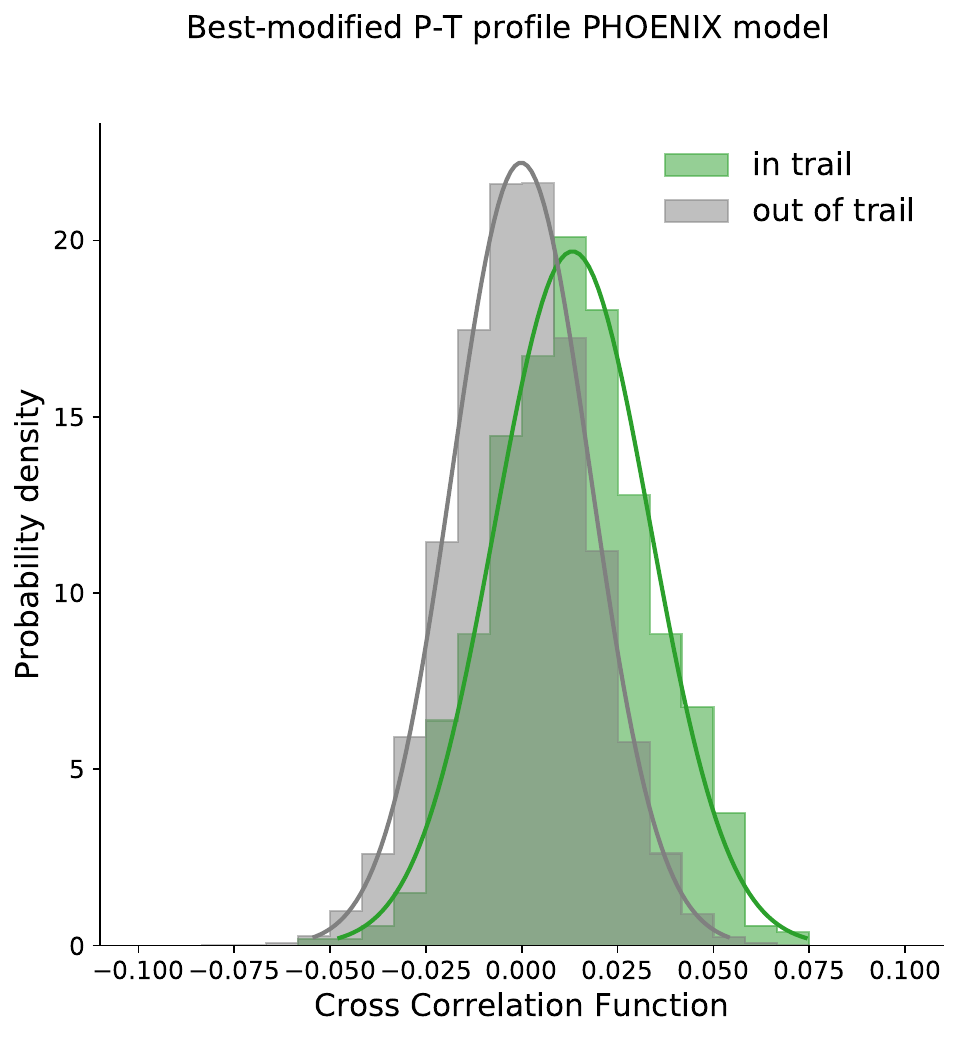}
    \caption{Histogram showing the probability density of the cross-correlation values in-trail (three most central columns) and out-of-trail (outside of the ten most central columns) of the exoplanet's velocity trail after aligning to the planet's rest frame. Values are shown for the PHOENIX best modified P-T profile model. The over-plotted solid lines are Gaussian fits to the distributions, which are well described by the Normal function.}
    \label{fig:wasp33_phoenix_best_modified_trails_hist}
\end{figure*}
\begin{figure*}
    \centering
    \includegraphics[width=\textwidth]{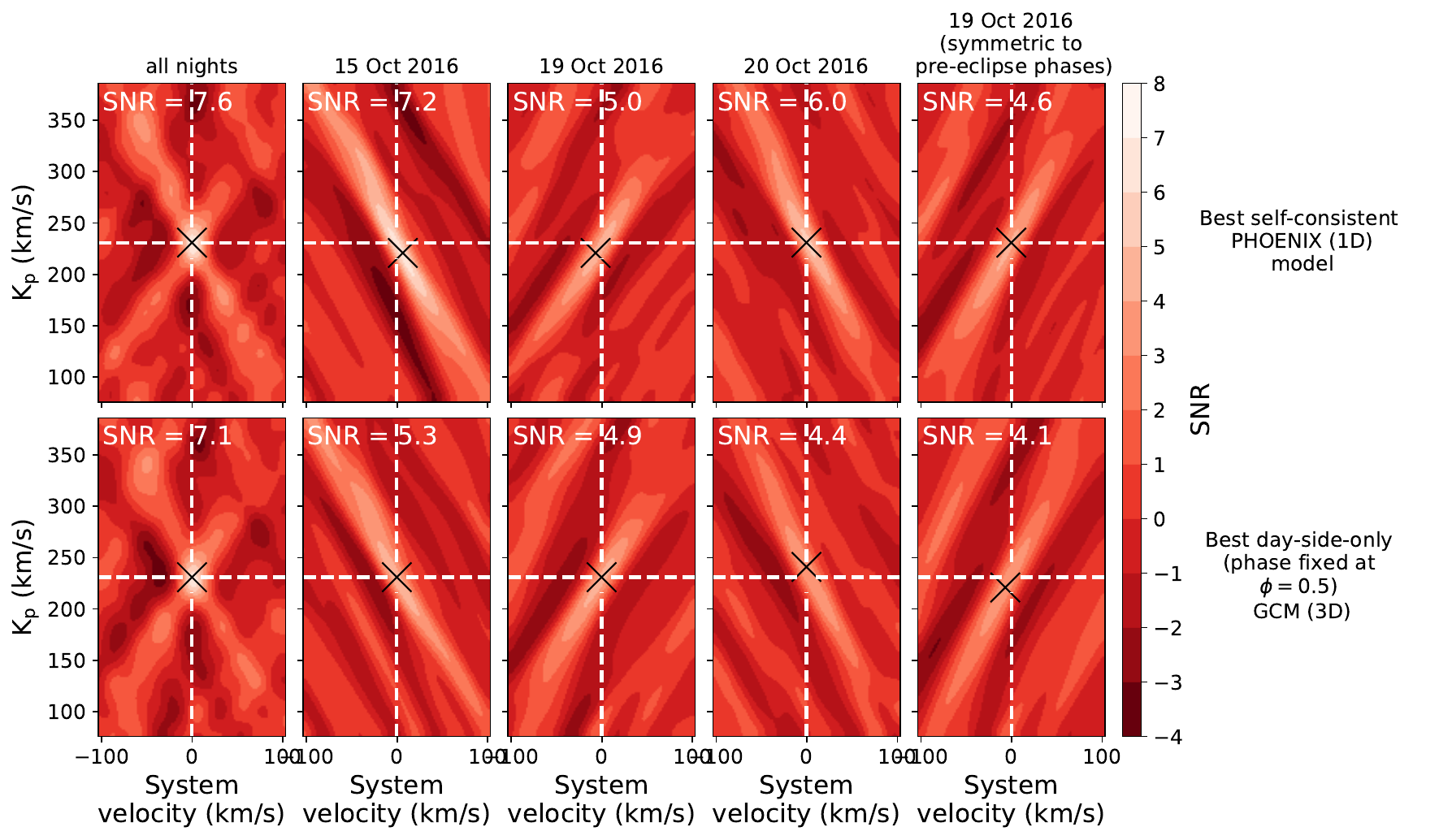}
    \caption{Same as Fig.~\ref{fig:snr_main} but shown for the best self-consistent PHOENIX model (top row), best day-side-only GCM (bottom row).}
    \label{fig:snr_appendix}
\end{figure*}
\begin{figure*}
    \centering
    \includegraphics[width=\textwidth]{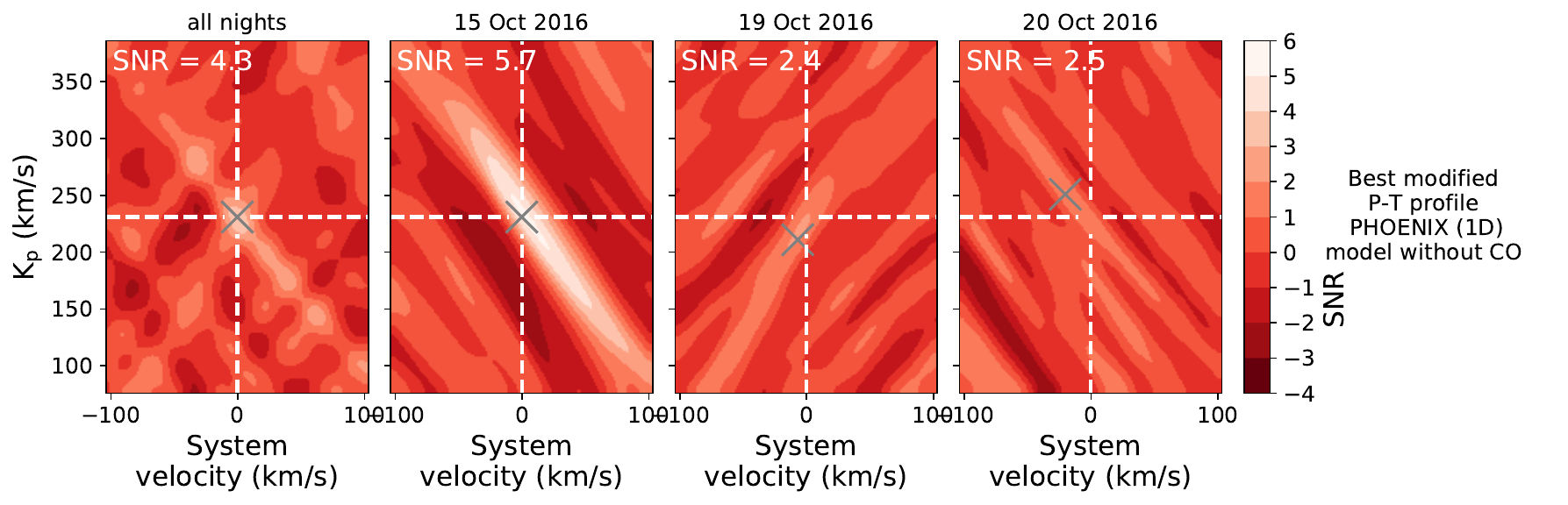}
    \caption{Similar to Fig.~\ref{fig:snr_main}, but for the best modified P-T profile PHOENIX model with all opacity sources included except CO. The detection of $5.7\sigma$ for the first night suggests we are still sensitive to some of the other opacity sources in the atmosphere of WASP-33 b.}
    \label{fig:snr_noCO}
\end{figure*}
\begin{figure*}
    \centering
    \includegraphics[width=\textwidth]{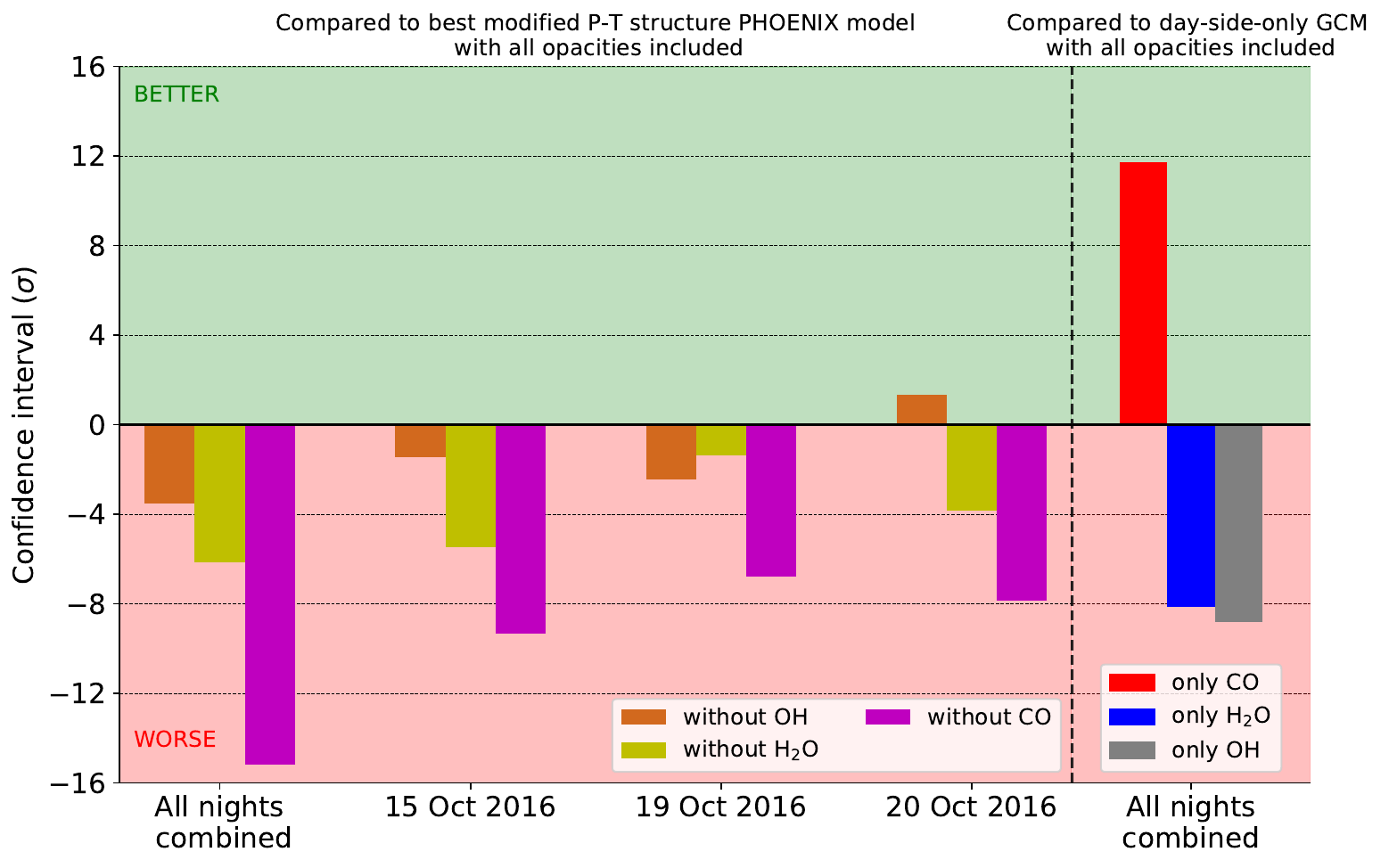}
    \caption{Comparison of the best modified P-T profile PHOENIX model (left-panel) and day-side-only GCM (right-panel) with all opacity sources included to models without/only certain opacity sources included. A negative confidence interval indicates the model is worse compared to the model where all opacity sources are included, whereas a positive confidence interval indicates the model is better. Confidence intervals are calculated using Wilks' Theorem.}
    \label{fig:opacities_wilks}
\end{figure*}
\begin{figure*}
    \centering
    \includegraphics[width=\textwidth]{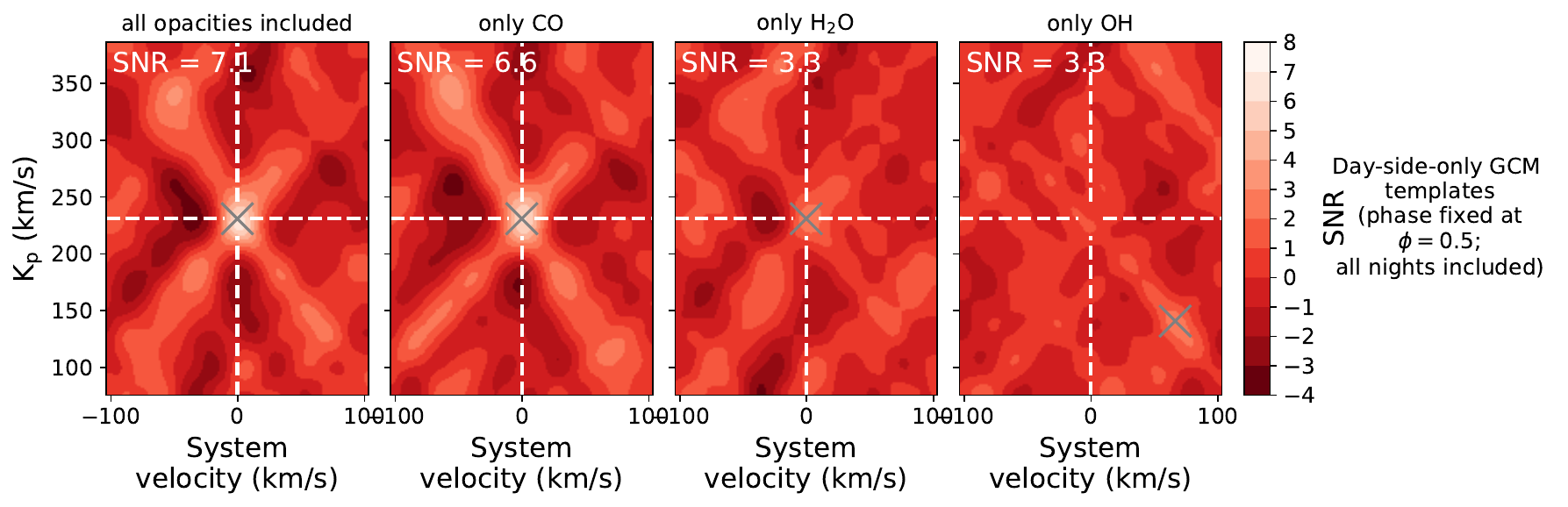}
    \caption{Similar to Fig.~\ref{fig:snr_main}, but for the GCM day-side-only templates with respectively: all opacity sources, only CO, only $\rm{H}_{2}O$ and only OH included. Amongst the opacity sources, CO is the only molecule that is robustly detected at 6.6$\sigma$.}
    \label{fig:gcm_species}
\end{figure*}
\begin{figure*}
    \centering
    \includegraphics[width=\columnwidth]{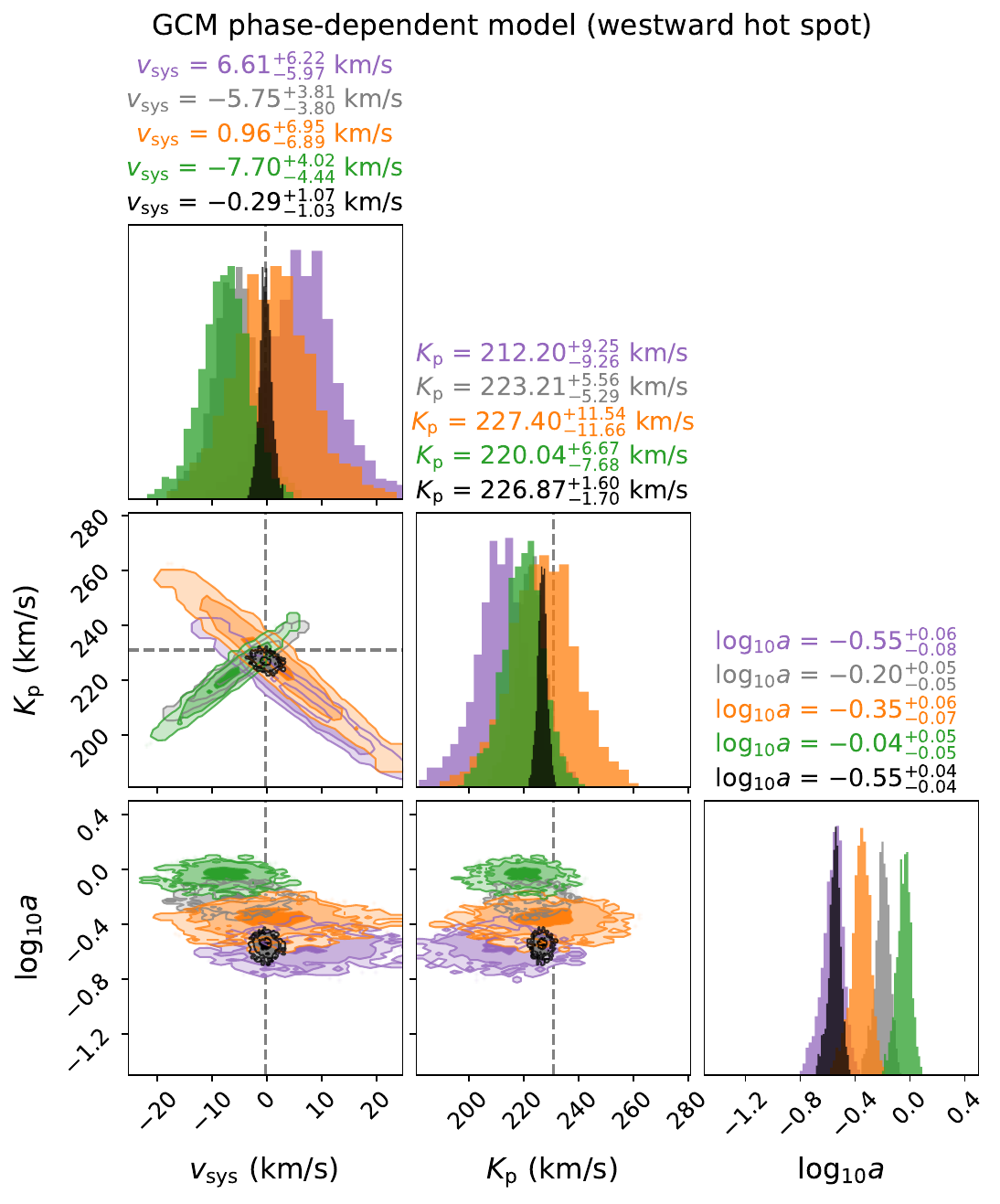}
    \caption{Same as the panels in Fig.~\ref{fig:corner_all}, but results shown for the GCM phase-dependent model where all phases have been mirrored such that eastward offsets become westward, effectively turning our model into one with a westward hot spot offset.}
    \label{fig:corner_westward}
\end{figure*}
\begin{figure*}
    \centering
    \includegraphics[width=\columnwidth]{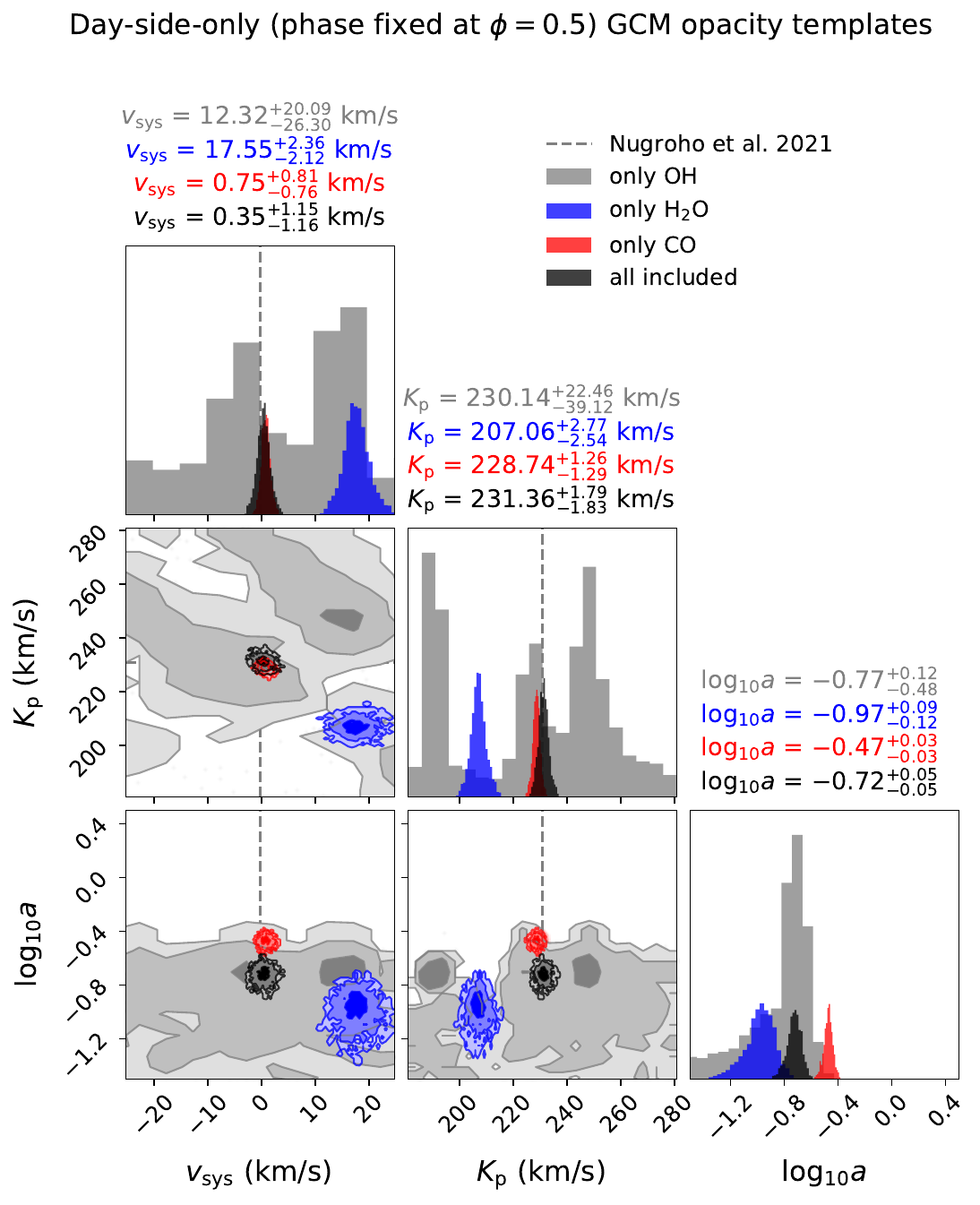}
    \caption{Same as the panels in Fig.~\ref{fig:corner_all}, but results shown for the GCM day-side-only spectral templates for different opacity sources included. All results are shown for all nights combined.}
    \label{fig:corner_opacity_models}
\end{figure*}


\bsp	
\label{lastpage}
\end{document}